\documentclass[twocolumn,prd,superscriptaddress,nofootinbib,preprintnumbers]{revtex4}

\usepackage{graphicx}
\usepackage{amsmath,bm,amssymb,amsfonts,dsfont}
\usepackage[usenames,dvipsnames]{xcolor}
\usepackage[normalem]{ulem}
\usepackage{url}
\usepackage{multirow}
\usepackage[colorlinks  = true,
            linkcolor   = NavyBlue,
            urlcolor    = NavyBlue,
            citecolor   = NavyBlue,
            anchorcolor = NavyBlue]{hyperref}

\newcommand{\diff}{\mathrm{d}}

\pdfoutput=1 

\newcommand{\lsim}{\mathrel{\mathop{\kern 0pt \rlap
  {\raise.2ex\hbox{$<$}}}
  \lower.9ex\hbox{\kern-.190em $\sim$}}}
\newcommand{\gsim}{\mathrel{\mathop{\kern 0pt \rlap
  {\raise.2ex\hbox{$>$}}}
  \lower.9ex\hbox{\kern-.190em $\sim$}}}

\newcommand{\KCh}{KoCu22}

\def \aap  {Astronomy \& Astrophysics}

\def \apj  {Astrophys. J.}

\def \jcap  {JCAP}
\def \prd {Phy. Rev. D}

\def \mnras {MNRAS}

\begin{document}

\preprint{LAPTH-014/23, TTK-23-07}

\title{A novel prediction for secondary positrons and electrons in the Galaxy}

\author{Mattia Di Mauro}
\email{dimauro.mattia@gmail.com}
\affiliation{Istituto Nazionale di Fisica Nucleare, via P. Giuria, 1, 10125 Torino, Italy}

\author{Fiorenza Donato}
\email{donato@to.infn.it}
\affiliation{Department of Physics, University of Torino, via P. Giuria, 1, 10125 Torino, Italy}
\affiliation{Istituto Nazionale di Fisica Nucleare, via P. Giuria, 1, 10125 Torino, Italy}

\author{Michael Korsmeier}
\email{michael.korsmeier@fysik.su.se}
\affiliation{Stockholm University and The Oskar Klein Centre for Cosmoparticle Physics, Alba Nova, 10691 Stockholm, Sweden}

\author{Silvia Manconi}
\email{manconi@lapth.cnrs.fr}
\affiliation{Laboratoire d'Annecy-le-Vieux de
Physique Théorique (LAPTh), CNRS, USMB, F-74940 Annecy, France}
\affiliation{Institute for Theoretical Particle Physics and Cosmology (TTK), RWTH Aachen University, D-52056 Aachen, Germany}

\author{Luca Orusa}
\email{luca.orusa@edu.unito.it}
\affiliation{Department of Physics, University of Torino, via P. Giuria, 1, 10125 Torino, Italy}
\affiliation{Istituto Nazionale di Fisica Nucleare, via P. Giuria, 1, 10125 Torino, Italy}

\begin{abstract}

The Galactic flux of cosmic-ray (CR) positrons in the GeV to TeV energy range is very likely due to different Galactic components. One of these is the inelastic scattering of CR nuclei with the atoms of the interstellar medium. The precise amount of this component determines the eventual contribution from other sources. We present here a new estimation of the secondary CR positron flux by incorporating the latest results for the production cross sections of $e^\pm$ from hadronic scatterings calibrated on collider data. 
All the reactions for CR nuclei up to silicon  scattering on both hydrogen and helium are included. The propagation models are derived consistently by fits on primary and secondary CR nuclei data. Models with a small halo size ($L \leq 2$ kpc) are disfavored by the nuclei data although the current uncertainties on the beryllium nuclear cross sections may impact this result. The resulting positron flux shows a strong dependence on the Galactic halo size, increasing up to factor 1.5 moving $L$ from 8 to 2 kpc.
Within the most reliable propagation models, the positron flux matches the data for energies below 1 GeV.
We verify that secondary positrons contribute less than $70\%$ of the data  above a few GeV, corroborating that an excess of positrons is already present at very low energies. At larger energies, our predictions are below the data with  the discrepancy becoming more and more pronounced.
Our results  are provided together with uncertainties due to propagation and hadronic cross sections. 
The former uncertainties are below 5\% at fixed $L$, while the latter are about 7\% almost independently of the propagation scheme. 
In addition to the predictions of positrons, we provide new predictions also for the secondary CR electron flux. 
\end{abstract}

\maketitle

\section{Introduction}
\label{sec:intro}

A guaranteed component of cosmic rays (CRs) is due to the so-called secondary 
production, originating from spallation reactions of CR nuclei against the atoms of the interstellar medium (ISM). Most of the secondary contribution is produced by the collision of CR protons or alpha particles interacting with hydrogen and helium ISM atoms. 
The secondary component plays an undisputed role in explaining the data collected by different space-based and ground-based experiments. 
This is particularly true for the fluxes of cosmic antiprotons 
\cite{Adriani:2012paa,PhysRevLett.117.091103} and positrons ($e^+$), which have been measured with high accuracy and in a wide energy range~\cite{Ackermann_2012,Adriani:2013uda,Adriani_2018,PhysRevLett.122.041102}. 
Indeed, the antiproton flux is explained at a large extent to be of secondary origin \cite{Donato:2008jk,Boudaud:2019efq}. 
On the other side, the measured $e^+$ flux and $e^+$ fraction, defined as the ratio between the flux of $e^+$ and the sum of $e^+$ and electrons ($e^-$), clearly indicate that a secondary component alone cannot explain the data \cite{Hooper:2008kg,Delahaye_2009,PhysRevLett.122.041102,Diesing:2020jtm}. 
In fact, secondary $e^+$ contribute mostly at energies below tens of GeV while at higher energies this process contributes to the data very likely less than a few tens of $\%$.
This is even more pronounced in the $e^-$ flux data, which are mainly explained by the cumulative flux of $e^-$ accelerated by Galactic supernova remnants \cite{DiMauro:2014iia,Aguilar:2019ksn,DiMauro:2020cbn, Evoli_2021}.

The presence of one or more astrophysical primary sources in the  $e^+$  flux has stimulated a vivid activity, exploring lepton production from astrophysical sources like pulsars and supernova remnants
\cite{Hooper:2008kg,Ahlers:2009ae,Boudaud:2014dta,Boudaud:2016jvj,Manconi_2017,Manconi:2018azw,Fornieri_2020,Manconi:2020ipm,DiMauro:2019yvh,Orusa_2021,Evoli_2021,Diesing:2020jtm,Cholis_2018,Cholis:2021kqk,Mertsch:2020ldv,
2012A&A...544A..16T}, and particle dark matter annihilation or decay into antimatter \cite{Cirelli:2008id,Bergstrom:2013jra,DiMauro:2015jxa,Di_Mauro_2021}.
The room left to primary $e^+$'s is gauged by the exact amount of secondary $e^+$'s 
predicted in the whole energy range in which data are available. One should notice that the most recent $e^+$ flux measurement by AMS-02 extends from 0.5 to 1000 GeV with an uncertainty smaller than $ 5\%$ for almost the whole energy range \cite{PhysRevLett.122.041102}.
Even though it is currently not achievable, significant effort should be devoted to produce a prediction of secondary production with a theoretical uncertainty that converges to the level of the AMS-02 data points. This is essential for investigating potential primary sources of $e^+$.

The flux of secondary $e^\pm$ is mainly determined by the physics of CR transport in the Galaxy, also known as CR propagation, and by the spallation and fragmentation cross sections of CRs scattering off the atoms of the ISM. A remarkable progress has been made on the propagation side, thanks to high quality data from AMS-02 nuclei and on parallel theoretical efforts to explain them 
\cite{Evoli:2017vim,Weinrich:2020ftb,DiMauro:2021qcf, Korsmeier:2021bkw, Korsmeier:2021brc,Maurin:2022gfm, Genolini:2021doh}. Nevertheless, the exact size of the Milky Way diffusive halo ($L$) is still not known. This has important consequences for the predictions of the flux of secondary cosmic particles (see, e.g., \cite{Weinrich:2020ftb}). 
Very recently, also the theoretical uncertainties on the parameterization of cross sections for the production of $e^\pm$ have been remarkably reduced thanks to a new determination of the Lorentz invariant cross section for the production of $\pi^\pm$ and $K^\pm$ by fitting data from collider experiments \cite{Orusa:2022pvp}. 
In that paper, the invariant
cross sections for several other channels contributing at the few \% level on the total cross section, as well as the contribution from scattering on nuclei, have been determined. The total differential cross section $d\sigma/dT_{e^\pm}(p+p\rightarrow e^\pm+X)$ was predicted from 10~MeV  up to 10~TeV of $e^\pm$ energy with an uncertainty of about 5-7\% in the energies relevant for AMS-02 $e^+$ flux. 
The result in \cite{Orusa:2022pvp} dramatically improved the precision of the theoretical model with respect to the state of the art \cite{Tan:1984ha,Blattnig:2000zf,Sjostrand:2014zea,Kelner:2006tc,Koldobskiy:2021nld,Kamae:2006bf,Kachelriess:2015wpa,Kachelriess:2019ifk,PhysRevD.15.820, Delahaye_2009,1998ApJ...493..694M}.

In this paper we provide a new evaluation of the CR flux of secondary $e^+$ and $e^-$ at Earth by implementing the new results on the production cross sections \cite{Orusa:2022pvp}.
In order to estimate the uncertainties coming from the propagation model, we perform a new fit to the 7 years fluxes of primary and secondary CRs measured by AMS-02 \cite{AGUILAR2020}, by using different assumptions for the physical processes that characterize the propagation of particles in the Galaxy and the diffusive halo size $L$. 
In particular, we estimate the uncertainties in the secondary flux which is due to various propagation parameters, devoting a specific discussion to the effect of the value of $L$, and to the $e^{\pm}$ production cross sections.
Our $e^+$ and $e^-$ secondary fluxes are predicted from the implementation of the innovative results from both sectors, production cross sections and Galactic propagation. 

The paper is organized as follows: In Sec.~\ref{sec:prod_prop} we summarize the modeling for the production and propagation of CRs in the Galaxy and we illustrate the benchmark models  used to compute the CR flux at Earth. In Sec.~\ref{sec:analysis} we explain our methods, by detailing how we solve the transport equation, and the strategies used to fit CR data  to calibrate the transport parameters. Our main results for the flux of secondary $e^{\pm}$, the primary and secondary CRs and the propagation parameters are discussed in Sec.~\ref{sec:results}. 
In Sec.~\ref{sec:conclusions} we draw our conclusions.
Finally, in the appendices we extend the discussion 
about the propagation parameters, the resulting fluxes of primary and secondary CRs, and further tests performed on the numerical solutions to the CR transport equation. 

\section{Cosmic-ray production and propagation}
\label{sec:prod_prop}

\subsection{The propagation of CRs in the Galaxy}
\label{sec:prop}

The charged particles injected in the ISM by their sources encounter several processes due to interaction with the Galactic magnetic fields, atoms or photons in the ISM, or Galactic winds.
All these processes can be modeled in a chain of coupled propagation equations for the densities $\psi_i$ of the CR species $i$. 
In general, $\psi_i$ depends on the position in the Galaxy ($\bm{x}$), the absolute value of the momentum ($p$), and time ($t$) (see, e.g., \cite{Strong:2007nh}):
\begin{eqnarray}
 \label{eq:propagationEquation}
 \frac{\partial \psi_i (\bm{x}, p, t)}{\partial t} = 
 q_i(\bm{x}, p) &+&  
 \bm{\nabla} \cdot \left( D_{xx} \bm{\nabla} \psi_i - \bm{V} \psi_i \right) \\  \nonumber
 &+& \frac{\partial}{\partial p} p^2 D_{pp} \frac{\partial}{\partial p} \frac{1}{p^2} \psi_i 
 \\  \nonumber
 &-& \frac{\partial}{\partial p} \left( \frac{\diff p}{\diff t} \psi_i  
 - \frac{p}{3} (\bm{\nabla \cdot V}) \psi_i \right)  \\  \nonumber
 &-& \frac{1}{\tau_{f,i}} \psi_i - \frac{1}{\tau_{r,i}} \psi_i.
\end{eqnarray} 
The equation content and the propagation setup is very similar to the ones discussed in \cite{Korsmeier:2021bkw}, \KCh~in the following.  Here we only explain the contents necessary to follow our analysis, and refer to \KCh\ for more details. 
From left to right, the equation describes eventual non stationary condition, the source terms, diffusion on the inhomogeneities of the Galactic magnetic field, convection due to the Galactic winds, reacceleration, energy losses, and catastrophic losses by fragmentations or radioactive decays. 
We model the diffusion coefficient by a double broken power law in rigidity, $R$, with the functional form
\begin{eqnarray}
    \label{eqn:diffusion_coefficient}
     D_{xx}(R) \propto \beta R^{\delta_l}
 	 \!\!&\cdot&\!\! \left( 1 + \left(\frac{R}{R_{D,0}}\right)^{\frac{1}{s_{D,0}}} \right)^{s_{D,0}\,( \delta - \delta_l) } \cdot \\ \nonumber 
 	\!\!&\cdot&\!\! \left( 1 + \left(\frac{R}{R_{D,1}}\right)^{\frac{1}{s_{D,1}}} \right)^{s_{D,1}\,( \delta_h - \delta) }.
    \end{eqnarray}
Here $\beta$ is the CR velocity in units of speed of light, $R_{D,0}$ and $R_{D,1}$ are the rigidities of the two breaks, $\delta_l$, $\delta$, and $\delta_h$ are the power-law index below, between, and above the breaks, respectively. We also allow a smoothing of the breaks through the parameters $s_{D,0}$ and $s_{D,1}$. 
The diffusion coefficient is normalized to a value $D_0$ at a reference rigidity of 4 GV so that $D_{xx}(R=4\,\rm{GV})=D_0$.
The first break, if included in the model, is typically in the range of 1–10 GV while the second break, whose existence is suggested by the flux data for secondary CRs, is at about 200-400 GV (see, e.g., \cite{Weinrich:2020cmw,Korsmeier:2021bkw,Korsmeier:2021brc,Vecchi:2022mpj}). 

The term $\bm{\nabla}(\bm{V} \psi_i)$ accounts for convection of CRs. We assume that the convection velocity is orthogonal to the Galactic plane $\bm{V}(\bm{x})= {\rm sign}(z) v_{{\rm c}}(z)\,{\bm e}_z$. 

Diffusive reacceleration describes  diffusion in momentum space through $D_{pp} \sim v_\mathrm{A}^2/D_{xx}$, where $v_\mathrm{A}$ is the speed of Alfv\'en magnetic waves. 
Energy losses are included in the propagation equations through the term ${\diff p}/{\diff t}$. 
Nuclear CRs can also encounter fragmentation due to the interaction with ISM atoms or decay. These processes are taken into account by the respective fragmentation and decay times $\tau_{f,i}$ and $\tau_{r,i}$. 

The source term for each primary CR species accelerated by astrophysical sources can be factorized as $q(\bm{x}, p) = Q(R) \rho(\bm{x})$.
The energy spectrum $Q(R)$ is parametrized as a smoothly broken power-law in rigidity:
\begin{equation}
 \label{eqn:energy_spectrum}
 Q(R) = Q_{0} R^{\gamma_1} \left( 1 + \left(\frac{R}{R_\mathrm{inj}}\right)^{1/s_{\rm{inj}}} \right)^{s_{\rm{inj}}\,( \gamma_2 - \gamma_1) }  ,
\end{equation}
where $R_\mathrm{inj}$ is the break rigidity, and $\gamma_1$ and $\gamma_2$ are the two spectral indices above and below the break. The smoothing of the break is parameterized by $s_{\rm{inj}}$.
For the spatial distribution of sources  $\rho(\bm{x})$ we assume the one of supernova remnants reported in Ref.~\cite{Green:2015isa}.

By solving Eq.~\eqref{eq:propagationEquation} one finds the interstellar CR density, for example at the location of the solar system.
Finally, we include the effect of the solar wind on particles entering the heliosphere, with the so called solar modulation, using the force-field approximation \cite{Fisk:1976aw}, which is fully determined by the solar modulation potential $\varphi$.
In particular, for CRs with rigidities above 1 GV the force-field approximation reproduces with a good precision the solar modulation of $e^+$ and $e^-$, as demonstrated in Ref.~\cite{Kuhlen:2019hqb}.
A similar conclusion is obtained by using SOLARPROP \cite{Kappl:2015hxv}, a code that numerically solves the transport of CRs in the heliosphere.  
By using input parameters similar to the standard ones suggested within SOLARPROP, we obtained results that closely aligned with the force-field approximation, for both $e^+$ and $p$. In particular, for $p$ with a rigidity above 1-2 GV and $e^+$ with an energy above 0.5-1 GeV, the differences between the SOLARPROP models and the force-field approximation are within the uncertainty range of the AMS-02 data.

\subsection{Secondary source term for cosmic nuclei and $e^{\pm}$}\label{sec:seclept}
Secondary CRs such as $e^\pm$ are produced in the interaction and fragmentation of primary CRs with the atoms of the ISM. 
The source term of secondary CRs is generically given by the convolution of the primary fluxes, the ISM components and the fragmentation cross-sections. 
In particular, for $e^\pm$:
\begin{eqnarray}    
    \label{eq:source_term}
    q(T_{e^{\pm}},\bm{x})&=& \sum_{i,j} 4 \pi\, n_{\mathrm{ISM},j}(\bm{x}) \times \nonumber \\
    && \times \int dT_i \, \phi_i(T_i,\bm{x})\frac{d\sigma_{ij}}{d T_{e^{\pm}}}(T_i,T_{e^{\pm}}) \,,
\end{eqnarray}
where $T_{e^{\pm}}$ is the $e^{\pm}$ kinetic energy, $\phi_i$ is the CR flux at the kinetic energy $T_i$, $n_{\mathrm{ISM},j}$ is the number density of the ISM $j$-th atom, and $d\sigma_{ij}/d T_{e^{\pm}}$ is the energy-differential production cross section for the reaction $i+j\rightarrow e^{\pm} + X$.
We note that, in general, the source term depends on the position in the Galaxy because both the CR gas density and the CR flux are a function of the position.
The factor $4\pi$ corresponds to the angular integration of the isotropic CR flux. 
Almost the entire ISM ($99\%$) consists of hydrogen and helium atoms \cite{Ferriere:2001rg}. The main channels for the production of secondary $e^{\pm}$ are $p+p$, $p+$He, He$+p$ and He+He.  

The calculation of the secondary $e^{\pm}$ follows from the $e^{\pm}$ production cross sections recently published in Ref.~\cite{Orusa:2022pvp},  which include all the possible channels due to pions, kaons and hyperons, and take into account nuclei contribution both in the ISM and in the incoming CR fluxes. The implementation of these new cross sections is the main novelty of this paper. They have been obtained with a very small uncertainty bands, whose effect is discussed in the following of this paper. 
In order to provide a reliable prediction for the secondary $e^{\pm}$ flux at Earth, we embark here in a novel determination of the propagation models by fitting the chain of primary and secondary nuclei on AMS-02 data (see Sect. \ref{sec:models} and \ref{sec:analysis}). 
The properties of source term for secondary $e^{\pm}$  in Eq.~\eqref{eq:source_term} have been discussed extensively in Ref.~\cite{Orusa:2022pvp}, to which we refer to any further detail (see e.g.~their Fig.~13). We have verified that the source term obtained by our new determination of the propagation models differs from the one illustrated in Ref.~\cite{Orusa:2022pvp} at most by 10\% in all the energy range.

In principle, the equation to calculate the source term of secondary nuclei is the same as for electrons and positrons. However, the kinetic energy per nucleon is conserved which simplifies Eq.~\eqref{eq:source_term}.
The uncertainty of fragmentation cross sections severely affects the production of secondary nuclei as well. 
The level of precision of fragmentation cross sections is for many channels significantly worse compared to the AMS-02 CR flux measurements (see, e.g., \cite{PhysRevC.98.034611}). 
Uncertainties are very often at the level of $20–30$\%, or even more for those cases with very poor data, and they represent the main limiting factor for the interpretation of the AMS-02 data.
For example, the uncertainties for the production cross sections of beryllium and its isotopes prevent to constrain precisely the value of the size of the diffusive halo (see, e.g., \cite{Evoli:2019iih,Maurin:2022gfm}). 
In our analysis we allow some flexibility in the fragmentation cross sections in order to take into account the related uncertainties. This is the same procedure used in several other papers (see, e.g., \cite{Cuoco:2019kuu,Weinrich:2020ftb,Korsmeier:2021bkw,Korsmeier:2021brc,Genolini:2021doh}).
\\
In particular, in Ref.~\cite{Weinrich:2020cmw} they used Gaussian priors for the scale, normalization and slope uncertainties in the cross sections. They fitted the average and the width of the Gaussian priors finding that the CR data need a rescaling of about $\pm 10-30\%$ with a change of slope of about $\pm 0.10$. We are going to use this result to justify our assumptions for the priors in the fragmentation cross sections.

\subsection{Models for cosmic-ray propagation}
\label{sec:models}

We test the following models for the propagation of CRs.
\begin{itemize}
\item {\tt Conv $v_{0,c}$}: 
It contains convection with a fixed velocity $v_{0,c}$ orthogonal with respect to the Galactic plane: $v_{c}(z) = v_{0,c}$. The CR injection spectra are taken as simple power laws ($\gamma_1=\gamma_2$ in Eq.~\eqref{eqn:energy_spectrum}) with separate spectral indexes for proton ($\gamma_{\rm{p}}$), Helium ($\gamma_{\rm{He}}$) and CNO ($\gamma_{\rm{CNO}}$).  The fact that these CR species have different injection spectra has been extensively demonstrated, e.g., in Refs.~\cite{Korsmeier:2021bkw,Korsmeier:2021brc,Genolini:2019ewc,Genolini:2021doh}. 
The observed low and high-rigidity breaks in CR fluxes are reproduced by a double smooth broken power-law shape for the diffusion coefficient as reported in Eq.~\ref{eqn:diffusion_coefficient}. The free propagation parameters are thus the following: $\gamma_p$, $\gamma_{\rm{He}}-\gamma_p$, $\gamma_{\rm{CNO}}-\gamma_p$, the diffusion coefficient parameters $D_0$, $\delta_l$, $\delta$, $\delta_h$, $R_{D,0}$, $R_{D,1}$, $s_{D,0}$ and $s_{D,1}$, $v_{0,c}$ and the same solar modulation potential $\phi$ for all the CR species.

\item {\tt Conv $dv_c/dz$}: 
This model is very similar to 
{\tt Conv~$v_{0,c}$}, but instead of using a constant convection velocity $v_{0,c}$, here $v_c(z)$ increases linearly as function of $z$. The exact functional form is $v_c(z) = dv_c/dz \cdot | z | $, where $dv_c/dz$ replaces $v_{0,c}$ as the free parameter in the fit.

\item {\tt Reacc$_{0}$}: 
This model has no convection while the reacceleration is turned on and modulated through the Alfv\`en velocity $v_a$, which is a free parameter in the fit. The diffusion coefficient and the injection spectra are modeled as in {\tt Conv $v_{0,c}$}. 
As we will see in Sec.~\ref{sec:results} the best-fit value for $v_a$ is around 0~km/s, this is why the label of the model reports 0 as subscript.

\item {\tt Reacc$_{10}$}: 
This model is the same as the previous one except that the Alfv\`en velocity is fixed to 10~km/s. 

\item {\tt Reacc$_{30}$-Inj}: 
In this model we replace the low-rigidity break in the diffusion coefficient with a low-rigidity break in the injection CR spectra. Therefore, we model the injection spectra of CRs with separate spectral indexes for p ($\gamma_{1,p}$ and $\gamma_{2,p}$), He ($\gamma_{1,\rm{He}}$ and $\gamma_{2,\rm{He}}$) and CNO ($\gamma_{1,\rm{CNO}}$ and $\gamma_{2,\rm{CNO}}$), which have a common rigidity break $R_{\rm{inj}}$ and smoothing $s_{\rm{inj}}$. The diffusion coefficient is modeled with a single smooth broken power-law with free parameters: $D_0$, $\delta$, $\delta_h$, $R_{D,1}$ and $s_{D,1}$,  and we leave free $v_{a}$.
This model has 30 as a subscript because the best-fit value for $v_a$ is found at 30~km/s.

\end{itemize}

For all the above mentioned models we also leave free the abundance of primary CRs. Specifically, we leave free the abundance of proton and helium  using a renormalization factor with respect to the reference values used in \textsc{Galprop}. For this, we iteratively adjust the reference isotopic abundances in \textsc{Galprop} to ensure that the renormalizations converge to values close to 1.
\footnote{Technically, in \textsc{Galprop} the isotopic abundance of protons is not fixed by the $Q_0$ of the source terms in Eq.~\eqref{eqn:energy_spectrum} but rather indirectly by aposteriori choosing  a normalization of the proton flux. In our case, we use  $4.3 \times 10^{-9}$ cm$^{-2}$ sr$^{-1}$ s$^{-1}$ MeV$^{-1}$ at 100 GeV. For all other primary CRs, the isotopic abundance is then provided as the ratio with respect to protons (in units of $1.06 \times10^6$). We fix the $^{ 4}\mathrm{He}$ abundance to $9.65  \times 10^{ 4}$.}
We call these parameters $\rm{Ren}\, \rm{Abd}_p$ and $\rm{Ren}\, \rm{Abd}_{\rm{He}}$ for proton and helium, respectively. This procedure is equivalent of having as free parameters the normalization factors $Q_0$ of the source terms in Eq.~\eqref{eqn:energy_spectrum} and allows a fast profiling over the parameters (see Sec. \ref{sec:analysis}). For the heavier nuclei, we leave free to vary the isotopic abundance of carbon $\rm  ^{12}C$, nitrogen $\rm ^{14}N$ and oxygen $\rm  ^{16}O$, which are all of primary origin, through the parameters: $\rm{Abd}_{^{12}C}$, $\rm{Abd}_{^{14}N}$ and $\rm{Abd}_{^{16}O}$. We do not use renormalizations as for $p$ and He here because the isotopic abundances also affect the fluxes of the secondaries Li, Be, and B.


\section{Methods}
\label{sec:analysis}

In this section we illustrate the methods used to evaluate the CR flux of secondary $e^+$ and $e^-$ at Earth. Specifically, we discuss the numerical solution of the transport equation, and the statistical methods for the determination of the propagation parameters and injection spectrum of primary CRs, obtained by fitting AMS-02 nuclei data. We then detail how we evaluate the uncertainties in the secondary $e^+$ and $e^-$ prediction coming from the new propagation models and from the production cross sections.

\subsection{Modeling cosmic-ray propagation with \textsc{Galprop}}

We employ the \textsc{Galprop} code\footnote{http://galprop.stanford.edu/} \cite{Strong:1999sv,2009arXiv0907.0559S} to solve the CR propagation equation numerically.
\textsc{Galprop}  divides the Galaxy, which is assumed to be a cylinder, in a spatial grid with respect to Galactocentric coordinates. We use the 2D grid where $r$ is the distance from the Galactic center and $z$ is the distance from the plane. 
We assume the Galactic plane to be extended 20 kpc, while for the half-height of the diffusion halo $L$ we test  values from 0.5 to 8 kpc. 
We discuss in the Appendix further details on the grid for the numerical solution of CR propagation with \textsc{Galprop}.
We include in the calculation of secondary leptons CR nuclei up to silicon. 
We use the new version of \textsc{Galprop} v.~57~\cite{Porter:2021tlr} which includes new solvers for the propagation equation, the possibility of using non-uniform grids, improved implementation of the convection velocity, new source distributions and improved parameterisations for calculations of the cross sections.

Particularly relevant ingredients for the prediction of the secondary $e^\pm$ are the ISM gas density and the treatment of the energy losses. 
For the gas, we use the 2D default models implemented in \textsc{Galprop} \cite{Porter:2021tlr}. 
The numerical solution of the transport equation permits to include all relevant energy losses for $e^\pm$, additionally modeling its spatial dependence. We include 
synchrotron losses on the Galactic magnetic field and inverse Compton losses on the interstellar radiation fields (ISRFs),  which are the dominant losses for $e^\pm$ detected at energies larger than about  10~GeV, as well as adiabatic, bremsstrahlung and ionisation losses, which affect the prediction at few GeV. 
The ISRF model is the default \textsc{Galprop} model, which is consistent  with more recent estimates in the few kpc around the Earth \cite{Vernetto:2016alq}, where most of the secondary leptons are produced. 
The synchrotron energy losses are computed by assuming a simple exponential magnetic field. Specifically, we include a regular magnetic field in the Galactic disk and a random component modeled as  exponential functions as 
$B_{\rm reg, ran} = B_{0, \rm reg, ran} \cdot \exp{(-(r- r_{\odot})/r_{0, \rm reg, ran}) \cdot \exp(- z/z_{0, \rm reg, ran})}$, with $B_{0, \rm reg, ran} =4\mu$G, $z_{0, \rm reg}=4$~kpc, $z_{0, \rm ran}=2$~kpc, $r_{0, \rm reg}=13$~kpc and an infinite $r_{0, \rm ran}$.  This gives a local total magnetic field of $B_{\rm tot} =\sqrt{B_{reg}^2 + B_{ran}^2 }=5.65 \mu$G, which is compatible with what found with state-of-the-art magnetic field spatial models fitted to CR and multiwavelength emissions \cite{Orlando:2019vmq}.

We have implemented the following custom modifications in \textsc{Galprop}, which have been detailed in Ref. \cite{Korsmeier:2021brc}. Smoothly broken power laws with up to two breaks are considered both for the primary injection spectra and for the diffusion coefficient, see Eqs.~\eqref{eqn:diffusion_coefficient},\ref{eqn:energy_spectrum}. The possibility to adjust the injection spectrum individually for each CR species is included, as well as nuisance parameters to allow freedom in the default fragmentation cross-sections for the production of secondary CRs.
A new custom modification introduced for this paper is the inclusion of the $e^{\pm}$ production cross section following the recent in Ref.~\cite{Orusa:2022pvp} as detailed in Sec.~\ref{sec:seclept}.

\subsection{Fit to nuclei cosmic ray data}

\subsubsection{Dataset}

\label{sec:data}
We fit the latest data measured by the AMS-02 experiment after 7 years of data taking, from 2011 to 2018, \cite{AGUILAR2020}.
In particular we fit the absolute fluxes of protons, He, C, O, N, B/C, Be/C and Li/C.
The ratio of secondaries over primaries (B/C, Be/C and Li/C) are particularly relevant for fixing the propagation parameters, while the one of He, C, O, N to derive the injection spectra.
Moreover, in the ratio some systematic uncertainties cancel out with respect to the absolute flux of secondary CRs.
Since all the AMS-02 measurements considered have been measured for the same data-taking period, we adopt one unique Fisk potential for the all the species.

The AMS-02 data for the fluxes available for $R>1$~GV are complemented with the proton and helium data from Voyager \cite{stonevoyager2013} above 0.1~GeV/nuc.
The addition of Voyager data helps to calibrate the interstellar injection spectrum. 
We use Voyager data only above 0.1 GeV/nuc to avoid further complications which might arise at very low energies, like stochasticity effects due to local sources or the possible presence of a further low energy break in the spectra \cite{Phan:2021iht}. 

The total number of data points considered in the analysis is 552. Since the number of free parameters in the model is between 25 and 30, a good $\chi^2$ is expected to be of the order or below 500.

\subsubsection{Fitting procedure}
\label{sec:method}
The statistical analysis performed in this paper is similar to the one presented in \cite{Korsmeier:2018gcy,Korsmeier:2021bkw,Korsmeier:2021brc}. We recap in what follows the key points and the novelties introduced.

The main goal of the analysis is to find the parameters of the model by fitting CR absolute flux data or flux ratios between secondary and primary CRs.
To optimize the computation time, we rely on a hybrid strategy to explore the wide parameter space, comprising of up to about 30 free parameters, as done in Refs.~\cite{Korsmeier:2021bkw,Korsmeier:2021brc}. We use the \textsc{MultiNest}~\cite{Feroz:2008xx} algorithm to sample all parameters that depend on the evaluation of \textsc{Galprop}\footnote{We use a \textsc{MultiNest} setting with 400 live points, an enlargement factor of \textsc{efr=0.7}, and a stopping criterion of \textsc{tol=0.1}.}. As a result, we obtain the posterior distributions and the Bayesian evidence.
For the other parameters which do not need a new evaluation of \textsc{Galprop} (for example, the renormalization of secondary CRs), we profile over those parameters on-the-fly with respect to the likelihood of Eq.~\eqref{eqn::likelihood_CR} and directly pass the maximum value to \textsc{MultiNest}. The profiling is performed using \textsc{Minuit}~\cite{James:1975dr}.
The best-fit and errors as well as the uncertainty bands for the fluxes and parameters correlations will be given in the Bayesian framework. We will use the Bayesian evidence  and Bayes factors to compare the different propagation models. In contrast, to quantify the goodness of fit for each model we employ the reduced chi-square statistics.

The posterior probability for the parameter $\theta_i$ is given by
\begin{eqnarray}
p(\theta_i|D) =  \int d\theta_1...d\theta_{i-1}d\theta_{i+1}...d\theta_{n} \; 
\frac{p(D|{\bm\theta}) p({\bm\theta})}{Z} \, ,
\end{eqnarray}
where $p(D|{\bm \theta})=\mathcal{L}({\bm\theta})$ is the likelihood given the data D, $p({\bm\theta})$ is the prior, and $Z=\int d{\bm\theta} p(D|{\bm\theta}) p({\bm\theta})$ is the evidence. 
As the log-likelihood we use a chi-square ($\chi^2$) function to compare our CR model with the available data:
\begin{eqnarray}
 	\label{eqn::likelihood_CR} 
 \log\left({{\cal L} }(\boldsymbol\theta_{\rm CR}, \boldsymbol\theta_{\rm XS})\right) &=& 
 -\frac{1}{2} \chi^2 (\boldsymbol\theta_{\rm CR}, \boldsymbol\theta_{\rm XS}) = \\ \nonumber 
   &=&  
    -\frac{1}{2} \sum\limits_{s,i}
							\left(\frac{\phi^{}_{{ s},i} - \phi^{(\text{m})}_{s,i} 
							(\boldsymbol\theta_{\rm CR}, \boldsymbol\theta_{\rm XS})}
							{ \sigma_{s,i} }\right)^2 
							\, .
\end{eqnarray}
Here the sum is performed over the CR data sets $s$ and the rigidity or energy bins $i$, and $\phi^{}_{{ s},i}$ and $\phi^{(\text{m})}_{s,i}$ are the measured and modeled CR flux of species $s$ at the rigidity $R_i$, respectively. The errors of the fluxes, labeled as $\sigma_{s,i}$, include both statistical and systematic uncertainties added in quadrature. 
We note that systematic uncertainties of the AMS-02 flux data are expected to exhibit correlation in $R$.
Such correlations play only a marginal role on the inferred propagation parameters \cite{Korsmeier:2021brc}, while they can have an important impact for dark matter searches with CR antiprotons~\cite{Boudaud:2019efq,Heisig:2020nse,Cuoco:2019kuu}. 

When fitting the model to AMS-02 data, we distinguish between two type of free parameters. The parameters $\boldsymbol \theta_{\rm CR}$ in Eq.~\eqref{eqn::likelihood_CR} are connected to the physics of CR propagation as introduced in Sec.~\ref{sec:prod_prop}.
Instead, the $\boldsymbol \theta_{\rm XS}$ are related to uncertainties in the nuclear fragmentation cross sections which are considered in the fit as nuisance parameters. 
This strategy permits marginalizing over the uncertainties in the fragmentation and production cross-sections, as introduced in Refs.~\cite{Korsmeier:2021bkw,Korsmeier:2021brc}.
In particular, we parametrize the cross sections for the production of secondaries CRs with a re-normalization factor, which for boron production is labeled as $A_{\rm{XS}} \rightarrow$ Be, and a change of slope, which for B is $\delta_{\rm{XS}} \rightarrow $Be.  
In particular, we use priors for the re-normalization and change of slope of the nuclear cross sections values of $0.8,1.2$ for the former, which means a variation of $20\%$, and $-0.1,0.1$ for the latter.

We provide a summary of the fit parameters and priors for each model tested in Appendix~\ref{app:tables}.
We assume linear priors for all the parameters.

\subsection{Secondary lepton prediction}

The predictions for the secondary $e^+$ and $e^-$ is computed once the propagation parameters best-fit and uncertainties have been found by fitting CR data as explained above.
In particular, we take the local CR flux found by fitting the data and then we compute, within the same propagation setup, the secondary $e^{\pm}$ fluxes due to the collision of CRs with the atoms of the ISM as in Eq.~\eqref{eq:source_term}.  
For each propagation model, the mean and the  $1\sigma$ Bayesian uncertainty are computed. This represents the statistical uncertainty connected to the fit to the CR propagation parameters only. 
This procedure is repeated for each propagation model benchmark. 
Additional uncertainty coming from the  $e^{\pm}$ production cross section is considered separately as obtained in Ref.~\cite{Orusa:2022pvp} and summed in quadrature  to build  the final uncertainty bands for the predictions.
We believe this choice to be conservative enough, since the propagation and cross section uncertainties can be considered independent and Gaussian to a good approximation. 
Additional systematic uncertainties connected for example  to the size of the diffusion halo or to the choice of the propagation model are discussed separately, and are found to be the dominant ones.

\section{Results}
\label{sec:results}

In this section we report the results  for the prediction of secondary electrons and positrons (Sec.~\ref{sec:secresults}), primary and secondary nuclei fluxes (Sec.~\ref{sec:CRflux}) and for the propagation parameters (Sec.~\ref{sec:propresults}).

\begin{figure}[t]
\centering
\includegraphics[width=0.49\textwidth]
{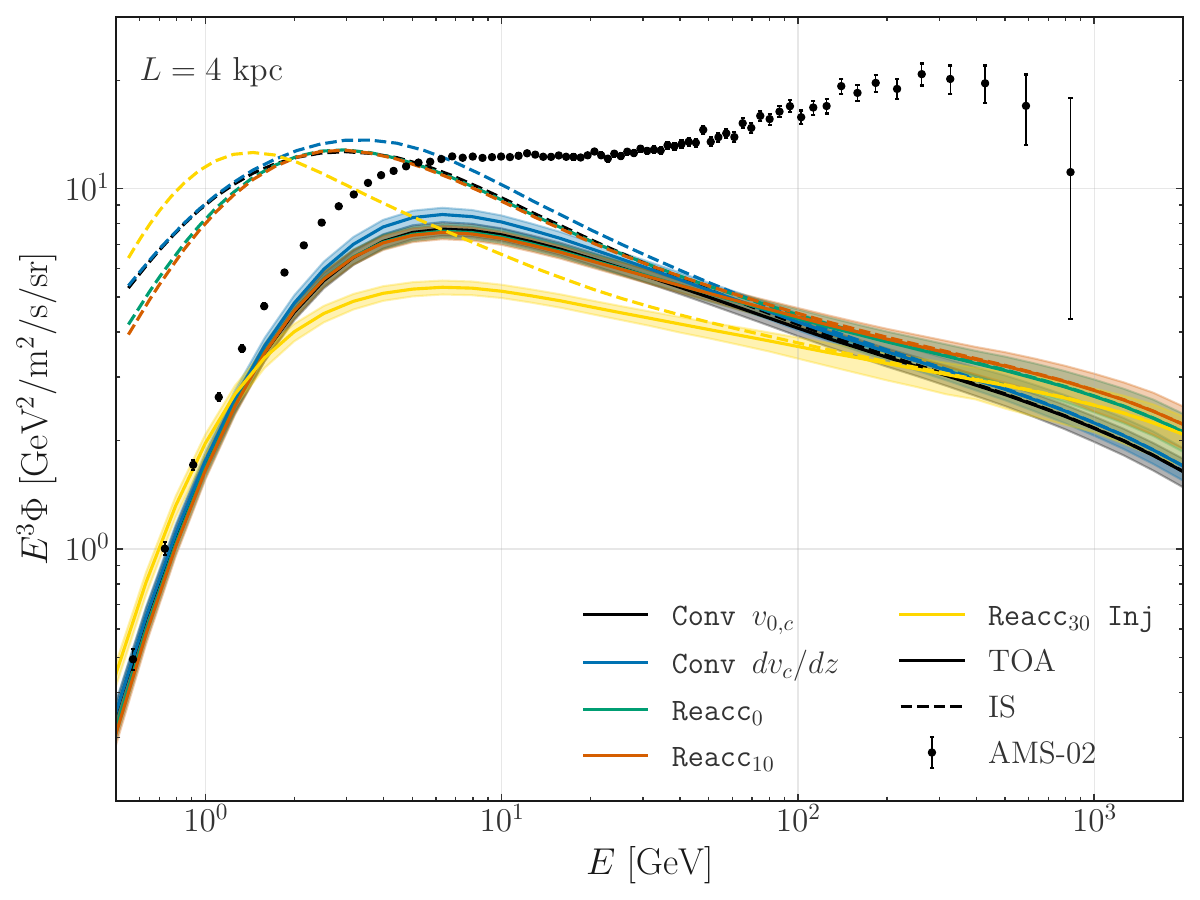}
\caption{
Prediction for the secondary positron flux at Earth as obtained for all the propagation model tested in this work (see Sec.~\ref{sec:models}) when fixing $L=4$ kpc. 
For each case, we show the interstellar (IS, dashed lines) and modulated top of atmosphere flux (TOA, solid lines). We display the best-fit and $1\sigma$ Bayesian
uncertainty band. AMS-02 data are included for comparison.}
\label{fig:sec_pos}
\end{figure}

\begin{figure}[t]
\centering
\includegraphics[width=0.49\textwidth]{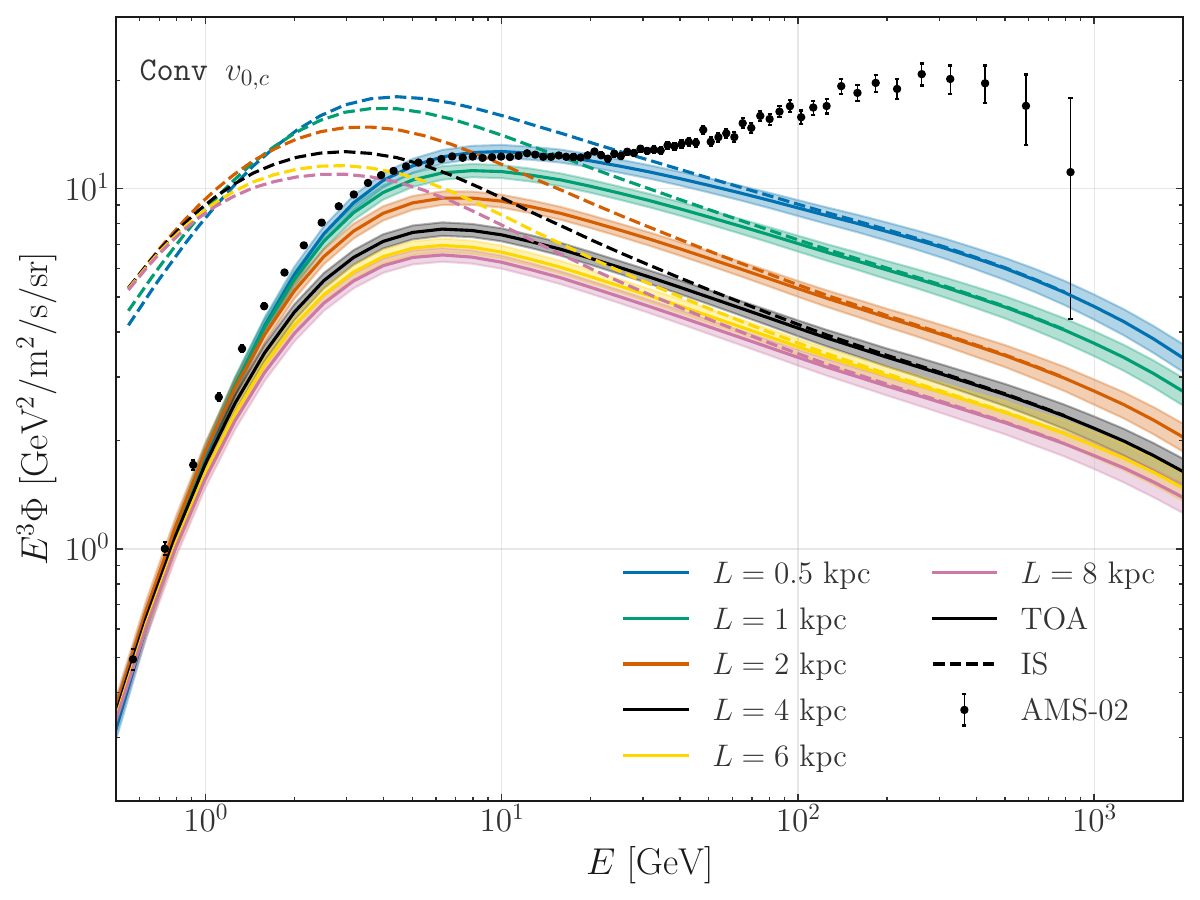}
\caption{
Prediction for the positron flux at Earth within the model {\tt Conv $v_{0,c}$} when varying the size of diffusive halo from $L=0.5$ to 8 kpc. Line styles and data as in Fig.~\ref{fig:sec_pos}. }
\label{fig:sec_pos_L}
\end{figure}

\subsection{Secondary leptons}
\label{sec:secresults}

In Fig.~\ref{fig:sec_pos} we display the predictions for the secondary positron flux obtained with all the different models introduced in Sec.~\ref{sec:models} for $L$ fixed to 4~kpc.
We show the best-fit and $1\sigma$ uncertainty band found in the Bayesian framework.
The models {\tt Conv $v_{0,c}$}, {\tt Reacc$_0$} and {\tt Reacc$_{10}$} predict a similar flux in the entire energy range. In particular, at the lowest measured energies the secondary fluxes are comparable to the $e^+$ data, while they are increasingly smaller with respect to the AMS-02 measurements at larger energies. At 5~GeV the secondary positrons can account for about 50-70$\%$ of the data while at the highest energy they are about 20-30$\%$ of the measured $e^+$ flux.

The {\tt Reacc Inj$_{30}$} provides a smaller flux by a factor of about 1.6 between 2 and 100~GeV with respect to the other cases, 
which can directly be related to the fact the model converges to a larger value for the diffusion coefficient. The larger diffusion coefficient in turn is partly obtained because we allow for larger uncertainties in the nuisance parameters of the nuclei fragmentation  cross sections, as further discussed in Sec. \ref{sec:propresults}. Moreover, this model is the only one that slightly overshoots the lowest AMS-02 data point at about 500 and 700 MeV. This result is expected because strong reacceleration significantly increases the lepton fluxes at low energies; similar results have been obtained by Ref.~\cite{Weinrich:2020ftb} in the QUAINT model. 

All models predict a similar flux of secondary $e^{+}$ at energies larger than 100 GeV, which is about a factor of five below the data. The variation at 1 TeV  is about a factor of two from the minimum to the maximum contribution.

In Fig.~\ref{fig:sec_pos_L} we show the $e^+$ flux predicted for different values of the diffusive halo size between 0.5 and 8 kpc within the {\tt Conv $v_{0,c}$} model. Above about 5 GeV, the secondary $e^+$ $E^3\,\Phi$ flux decreases systematically with $L$. This  can be understood from the well-known degeneracy between $L$ and the normalization of the diffusion coefficient \cite{Maurin:2001sj}. 
For small $L$, CR nuclei spend on average more time in the Galactic disc, which increases the secondary nuclei production. The latter has then 
to be compensated by a smaller diffusion coefficient (i.e. faster diffusion). Therefore, to a first approximation, CR nuclei data only constrain the ratio $D_0/L$. Indeed, we confirm in Sec.~\ref{sec:propresults} that there is a linear correlation between $L$ and $D_0$. In contrast, $e^+$ (and also $e^-$) suffer from stronger energy losses which restrict them more locally than nuclei, such that they do not perceive the same effect of the boundary at $L$ as nuclei. For them the degeneracy between $L$ and $D_0$ is broken and they only sense the effect of decreasing $D_0$, which increases the secondary flux. For $L=0.5$ kpc the $e^+$ flux is at the level of the data between 0.5 to 20~GeV, while the flux for $L=2$~kpc (4 kpc) decreases by of 20$\%$ (40$\%$) at 5 GeV. For 8 kpc, the predicted secondary flux is about $50\%$ of the data at 5 GeV. 
The predictions obtained with different $L$ converge to very similar values  below 2 GeV because energy losses become less important at small positron energies. The contribution of secondary positrons to the highest AMS-02 energy at $E\sim$~TeV spans from few percent to 50\% of the data, mostly depending on the value of $L$. 

\begin{figure}[t]
\centering
\includegraphics[width=0.49\textwidth]{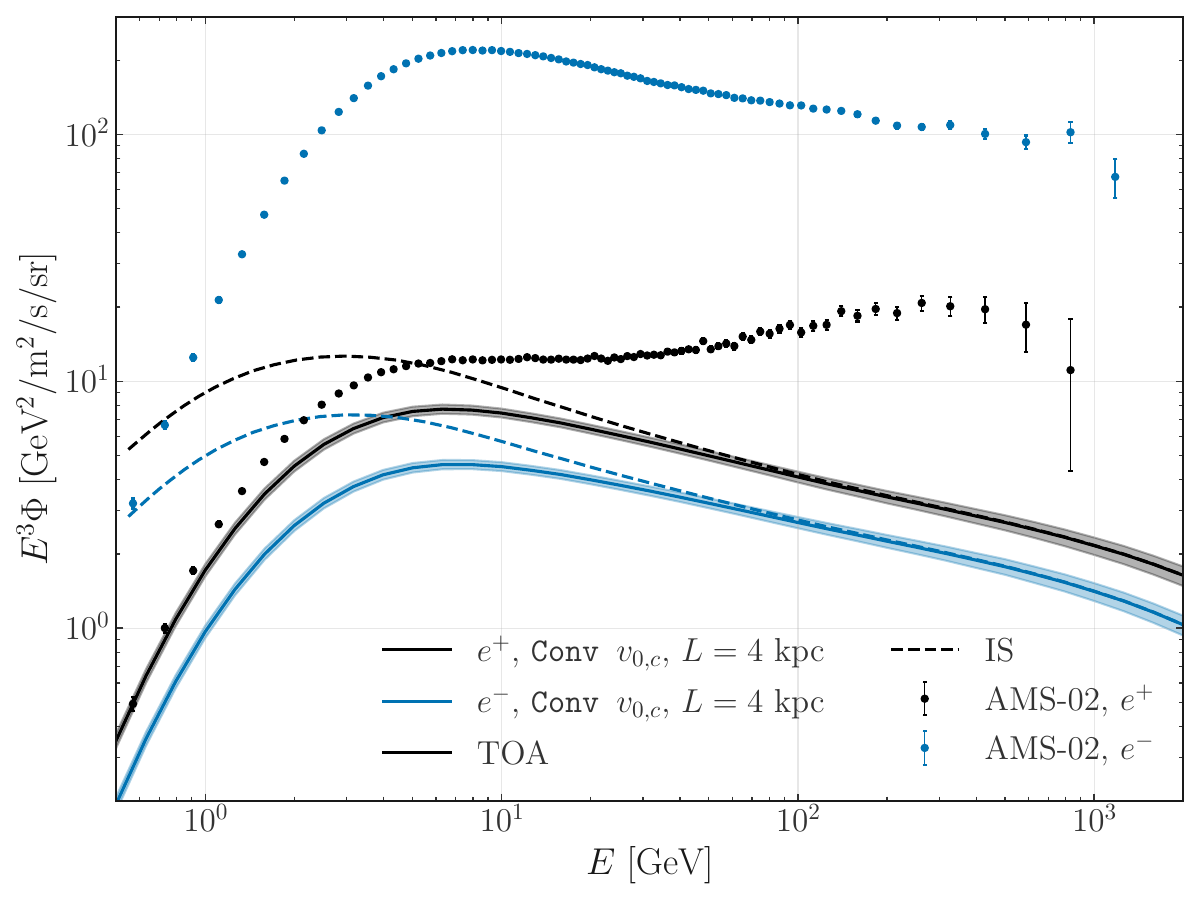}
\caption{Flux of positrons (black line and band) and electrons (blue line and band) obtained for the model {\tt Conv $v_{0,c}$} with $L=4$ kpc. We show the AMS-02 data for positrons (black data) and electrons (blue data).} \label{fig:sec_elepos}
\end{figure}

In Fig.~\ref{fig:sec_elepos} we show the flux for secondary electrons and positrons compared to $e^{\pm}$ AMS-02 data. As expected, secondary electrons have a smaller flux with respect to positrons, reflecting the charge asymmetry in the colliding CR and ISM particles, mostly positively charged.
We verified that the variation of the secondary electrons with the size of the diffusive halo and propagation model follows the $e^+$ trends, as shown in Fig.~\ref{fig:sec_pos} and Fig.~\ref{fig:sec_pos_L}.

\begin{figure}[t]
\includegraphics[width=0.5\textwidth]{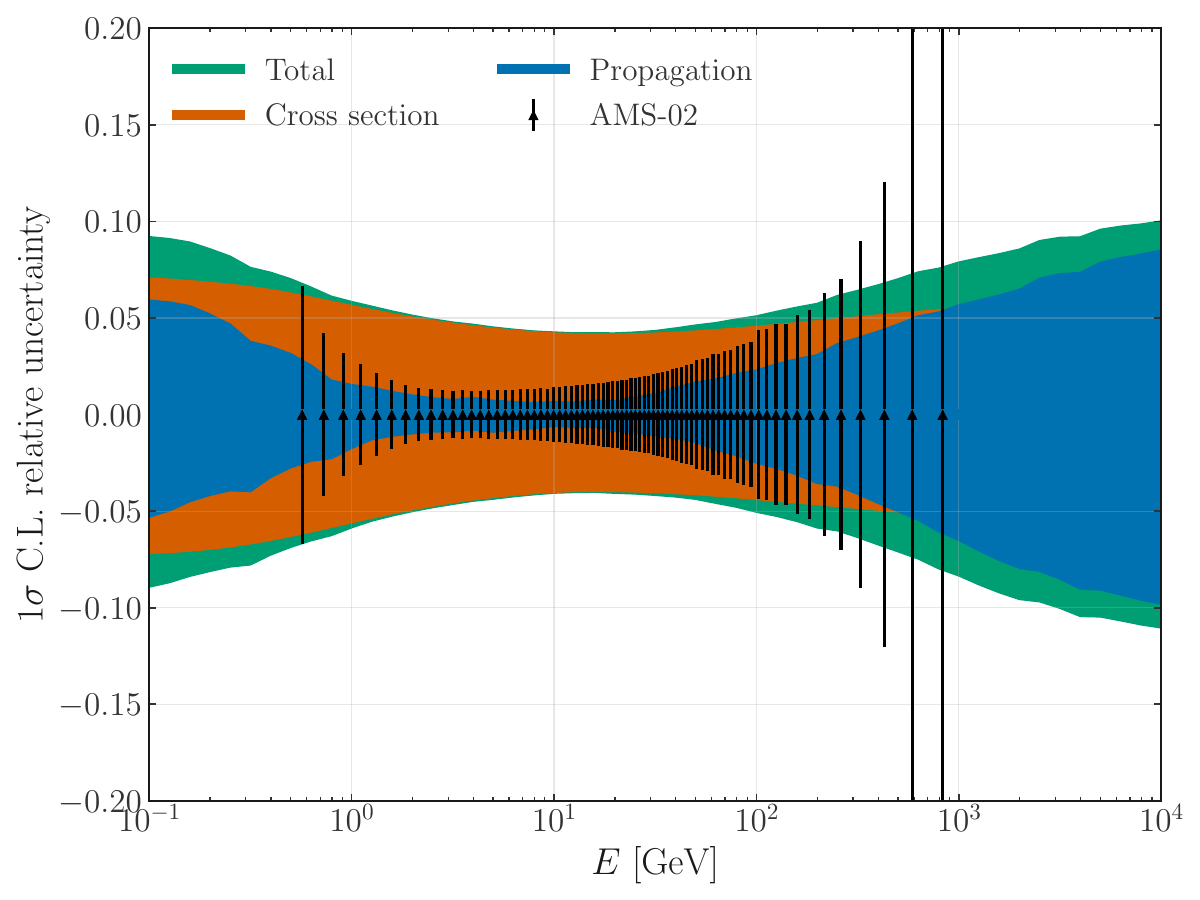}
\caption{Uncertainty for positron flux due to the propagation parameters and $e^{\pm}$ production cross sections for the case {\tt Conv $v_{0,c}$} with $L=4$ kpc. We also show the total uncertainty obtained with the sum in quadrature of the two uncertainties. For comparison, we show the errors of the $e^+$ AMS-02 data.} \label{fig:sec_elepos_unc}
\end{figure}

In all the predictions shown in Figs.~\ref{fig:sec_pos} -- \ref{fig:sec_elepos} we report the uncertainty band related to the fit to the CR propagation parameters and on the $e^\pm$ cross sections. 
We detail in Fig.~\ref{fig:sec_elepos_unc}  the uncertainties related to both contributions for the case {\tt Conv $v_{0,c}$} with $L=4$ kpc. The propagation parameters' uncertainties are in general smaller than the cross section ones up to 1 TeV, above which they both reach 10$\%$,
and they are at the level of  few \% between 1 and 100 GeV, always comparable or smaller than the size of experimental errors.  The latter are shown as the sum in quadrature of the AMS-02 statistical and systematic errors on the $e^+$ flux. 
Instead, the uncertainties related to the $e^{\pm}$ production cross sections are almost energy independent and at the level of $5-7\%$.

\subsection{Primary and secondary nuclei}
\label{sec:CRflux}

\begin{figure*}[t]
    \centering
    \includegraphics[width=0.49\textwidth]{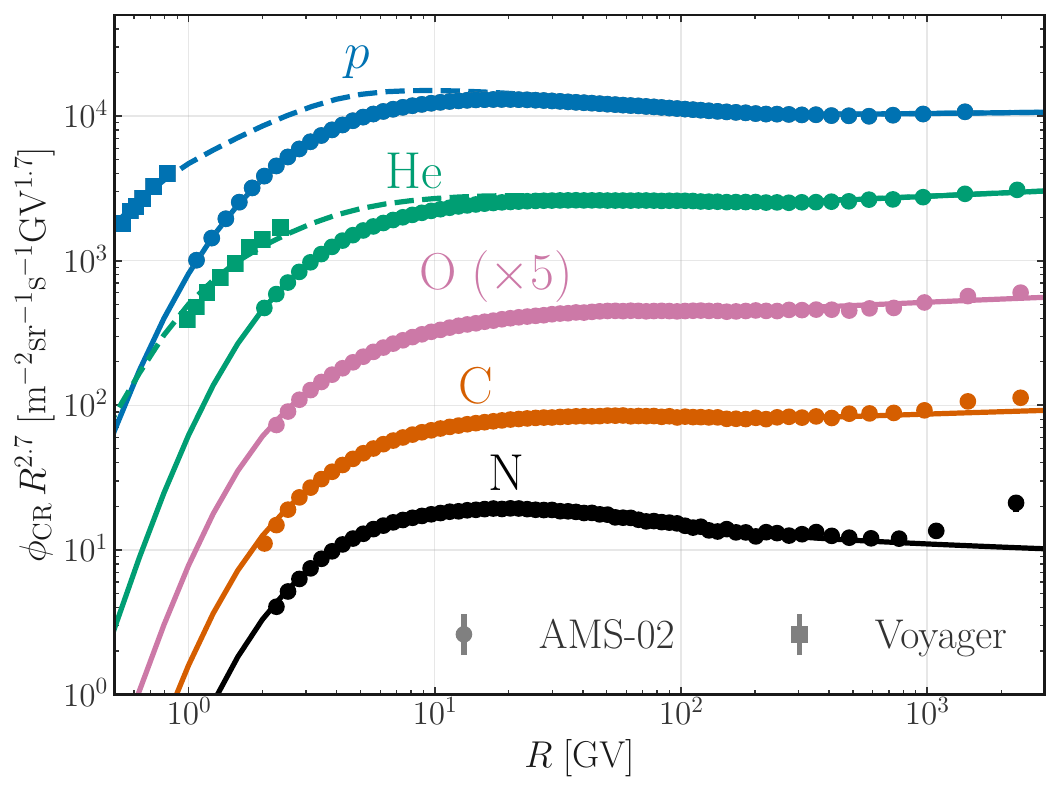}\includegraphics[width=0.49\textwidth]{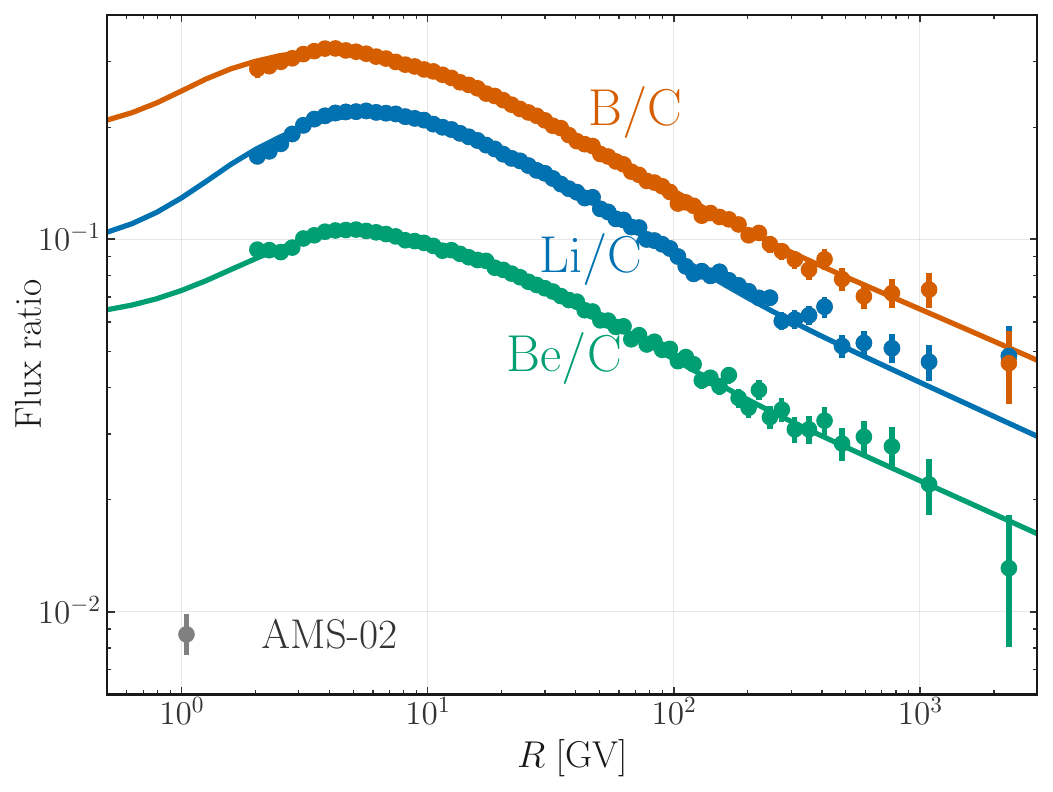}
    \caption{Left: Fluxes of p, He, O, C and N nuclei as predicted by the parameters fitted on the data by AMS-02 and Voyager (for p and He). Right: secondary-to-primary flux ration for B/C, Li/C and Be/C along with AMS-02 data.}
    \label{fig:nuclei}
\end{figure*}
We show here the results for the fit to the primary and secondary CR nuclei.
The procedure that we use for fitting the data is explained in Sec.~\ref{sec:analysis} while the free parameters in each model  are reported in Sec.~\ref{sec:models}.
Fig.~\ref{fig:nuclei} summarizes the results of the fit for {\tt Conv $v_{0,c}$} along with the AMS-02 data for primaries and secondaries species as a function of rigidity. In the left panel we report the results for the primary p, He, O and C nuclei, and the half-primary N flux. On the left panel we show the secondary-to-primary flux ratios for B/C, Li/C and Be/C. 
In appendix \ref{app:tables} we report the best-fit values for the propagation parameters and the residual plots for different cases. 

We obtain reduced $\chi^2$ ($\tilde{\chi}$) smaller than 1 within each of the tested models. 
The convective models {\tt Conv $v_{0,c}$} and {\tt Conv $dv/dz$} have $\tilde{\chi}\approx0.7-0.8$ and the models {\tt Reacc$_0$} and {\tt Reacc$_{10}$} $\tilde{\chi}\approx0.8-0.9$.
The lowest $\tilde{\chi}$ of $0.50$ is provided by the {\tt Reacc$_{30}$ Inj} model.
The Bayesian evidence is $\log(Z)=-207.9$ for the {\tt Reacc$_{30}$ Inj} and $\log(Z)=-237.7$ for the {\tt Conv $v_{0,c}$} model, which implies a statistically strong preference of the first model.
We note, however, that this is the only model for which we allow for larger priors of the nuclear cross section uncertainties and with the highest number of  free propagation parameters. 
Namely, instead of only 3 free slopes in the {\tt Conv $v_{0,c}$}, this model  has 6 free slopes as well as a free position and smoothing of the break. We confirmed that indeed the improvement of $\tilde{\chi}$ mostly comes from the primary CR spectra. 

The fact that all our models converge to a best-fit with of $\tilde \chi$ smaller than one is expected and in agreement with previous studies. The reason for the small $\tilde\chi$ values is that the systematic uncertainties of the CR data points of AMS-02 are correlated. Those correlations are not provided by the collaboration. However, there have been different attempts to model these correlations \cite{Boudaud:2019efq,Heisig:2020nse,Cuoco:2019kuu}. The correlations typically slightly reduce the uncertainty on the propagation parameters and increase the $\tilde \chi$. Taking them into account can be crucial, for example, when searching for dark matter signatures in CR antiprotons. 
In the absence of correlations provided by the AMS-02 collaboration we follow a conservative approach and assume uncorrelated uncertainties, adding statistical and systematic uncertainty in quadrature for each data point as explained above.

In contrast, the quality of the fit to p and He Voyager data is slightly worse with a $\tilde{\chi}=2-3$. However, the Voyager data are at very low energy, below the main focus of this work. A better fit of those data typically would require an additional low-energy break in the injection spectra \cite{Vittino:2019yme}.
Our models, except for {\tt Reacc$_{30}$ Inj}, does not perfectly fit the highest energy data points, especially the N spectra, see Appendix \ref{app:tables}. 

\subsection{Propagation and cross section parameters}
\label{sec:propresults}
In this section we report the results on the propagation parameters as derived from the fits to the nuclear data.
We start by discussing the results of the {\tt Conv $v_{0,c}$} model for different values of $L$.
The injection spectra of primaries are well constrained. For $L\geq 2$ kpc, the injection slope for protons is very similar and converges to values between 2.36 and 2.37, while for smaller $L$ the spectrum softens slightly to $\gamma_p$ up to 2.40. We find that the injection slope for He and CNO are significantly different from proton by about 0.055 and 0.02, respectively. 

\begin{figure}
    \centering
    \includegraphics[width=0.49\textwidth]{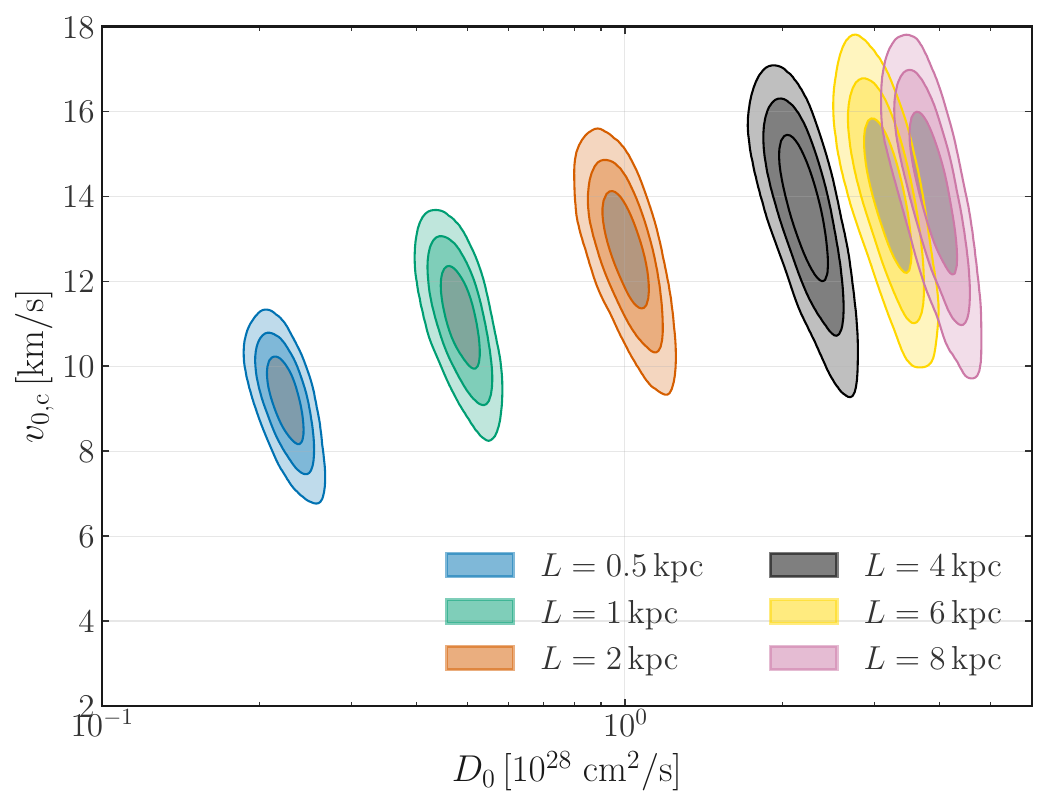}
    \caption{This figure shows the $1,2,3\sigma$ Bayesian contours for the parameters $D_0$ and $v_{0,c}$ obtained from the fit to the CR data when we assume different $L$.}
    \label{fig:D0v0}
\end{figure}

In contrast, the diffusion coefficient  changes significantly as function of $L$. The main impact concerns its normalization $D_0$. This is due to the well-known degeneracy between $L$ and $D_0$ \cite{Maurin:2001sj} as already discussed above. By fitting a power law to the fit's result for $L\in[0.5,6]$~kpc we obtain the empirical relation:
\begin{equation}
 D_0(L) = 4.70 \cdot 10^{27} \, \rm{cm}^2/\rm{s} \, \left(\frac{L}{\rm{kpc}}\right)^{1.08\pm0.01}\, .
\label{eqn:LD}
\end{equation}
The slope of $1.08\pm0.01$ is close to 1 indicating that a $L$ and $D_0$ are almost direct proportional to each other. We note that for large $L$ the relation breaks down because the height of the Galactic halo starts to be comparable to the radial size of the Galaxy. We see that already at $L=8$~kpc this relation starts to break, explaining why we did not include it in the fit of Eq.~\eqref{eqn:LD}.
Next to the strong correlation of $D_0$ and $L$ there is a smaller correlation between $v_{0,c}$ and $L$. 
We show in Fig.~\ref{fig:D0v0} the $1,\ 2,$ and $3\sigma$ contours, obtained in the Bayesian framework for the parameters $D_0$ and $v_{0,c}$ obtained from the fit to the data and assuming different $L$ sizes.
The best-fit values of $v_{0,c}$ increase as a function of $L$, namely, we find 9 km/s for $L=0.5$ kpc and 14 km/s for $L=8$ kpc.
Moreover, for fixed $L$ there is a small anticorrelation between $D_0$ and $v_{0,c}$ meaning that for smaller values of $v_c$ it is possible to have larger values of $D_0$.

\begin{figure*}
\includegraphics[width=0.49\textwidth]{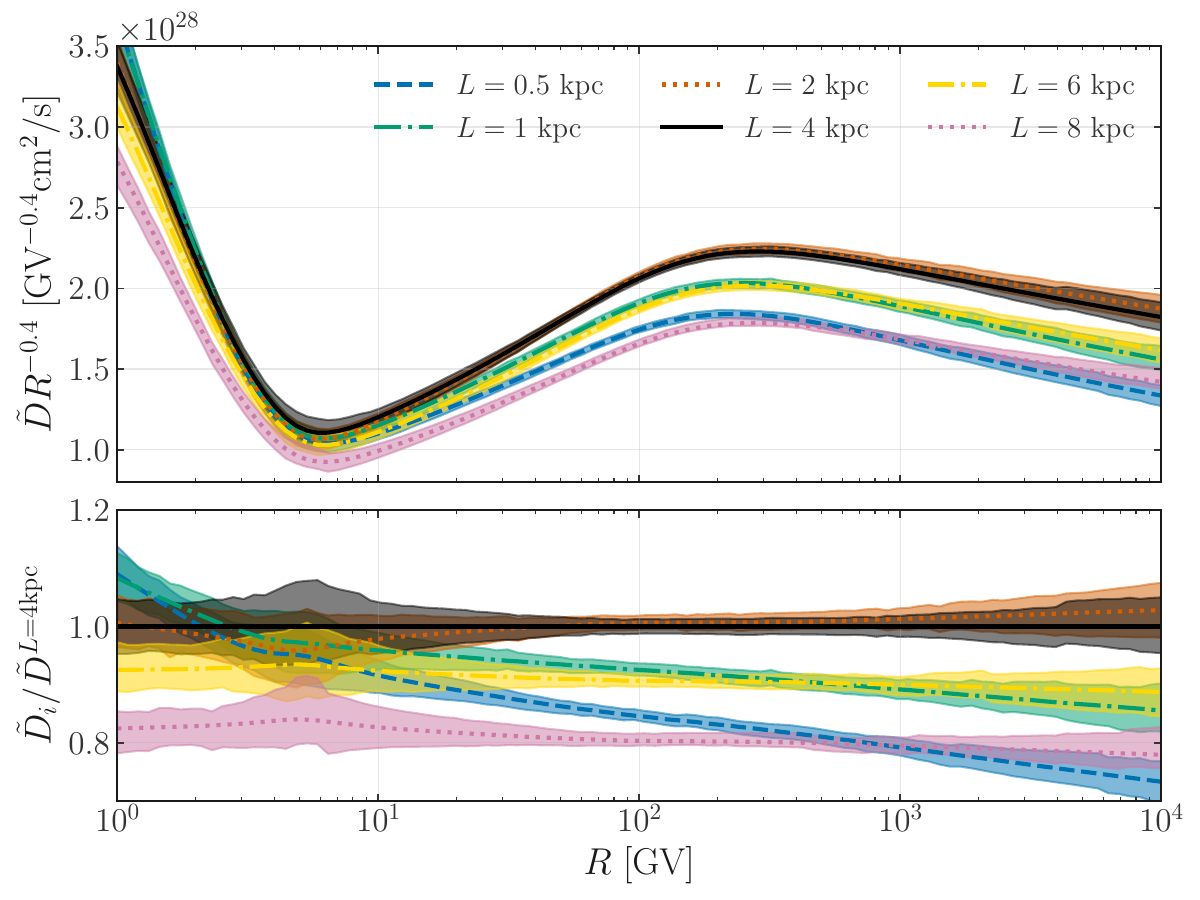}
\includegraphics[width=0.49\textwidth]{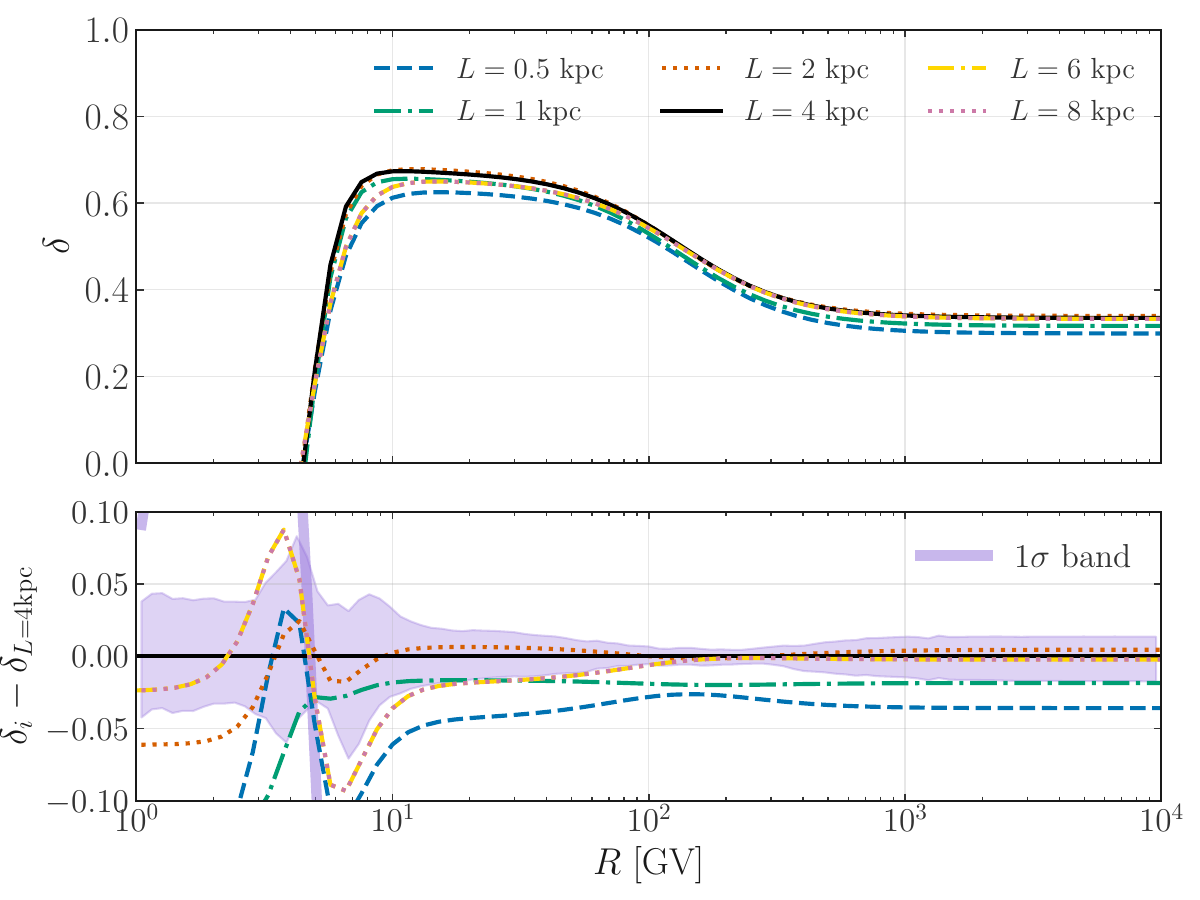}
\caption{Left Panel: Shape of the diffusion coefficient rescaled as $\tilde{D}(R)R^{-0.4}$ as a function of $R$ found from the fit to the nuclei data for different values of $L$. The value of $D_0$ is rescaled according to  Eq.~\ref{eq:rescD0} at $L=4$ kpc. The bottom part of the plot shows the ratio between the value obtained in the different cases with respect to the one obtained in the case where $L=4$ kpc. The bands represent the $1\sigma$ C.L. band obtained from the fit to CR data in each case. 
Right panel: slope of the diffusion coefficient $\delta(R)$ as a function of rigidity, as selected by the fit to the nuclei data. The bottom part of the plot shows the difference between the value of $\delta$ obtained in the different cases with respect to the one obtained in the case where $L=4$ kpc. The purple band represents the $1\sigma$ C.L. band obtained from the fit to CR data.} 
\label{fig:diffusionslopes}
\end{figure*}

The shape of the diffusion coefficient as a function of rigidity is very similar for all $L$, as we show in Fig.~\ref{fig:diffusionslopes}.

In order to focus on the shape rather than the normalization, we use Eq.~\eqref{eqn:LD} to rescale all the diffusion coefficients to $L=4$~kpc, more specifically, we define the rescaled diffusion coefficient:
\begin{equation}
\label{eq:rescD0}
\tilde D = (4\,\mathrm{kpc}/L)^{1.08} D\,.
\end{equation}

All the curves in the left panel of Fig.~\ref{fig:diffusionslopes}, except the case with $L=8$ kpc, which is not fitted, have the same normalization within the $1\sigma$ error band at 4 GV, where the value of $D_0$ is fixed. The differences at lower and higher rigidities are due to small differences in the best-fit values of the slope parameters.

The case for $L=8$ kpc has a $20\%$ normalization shift with respect to the other cases because, as explained before, the correlation between $D_0$ and $L$ breaks for large values of L. 

In order to compare the shape of $D(R)$ we plot in the right panel of Fig.~\ref{fig:diffusionslopes} the slope of that function defined as $\delta(R) = dD/dR$.
As clearly shown in the figure, the slope of the diffusion coefficient is very similar in all the tested cases. In particular the variations obtained for the best cases in the paper, i.e.~for $L>1$ kpc, are compatible within the statistical errors.
In addition, the shape of $\delta(R)$ that we find here is similar with respect to the results in Refs.~\cite{Weinrich:2020cmw,Maurin:2022gfm}. 

\begin{figure}[t]
    \includegraphics[width=0.485\textwidth]{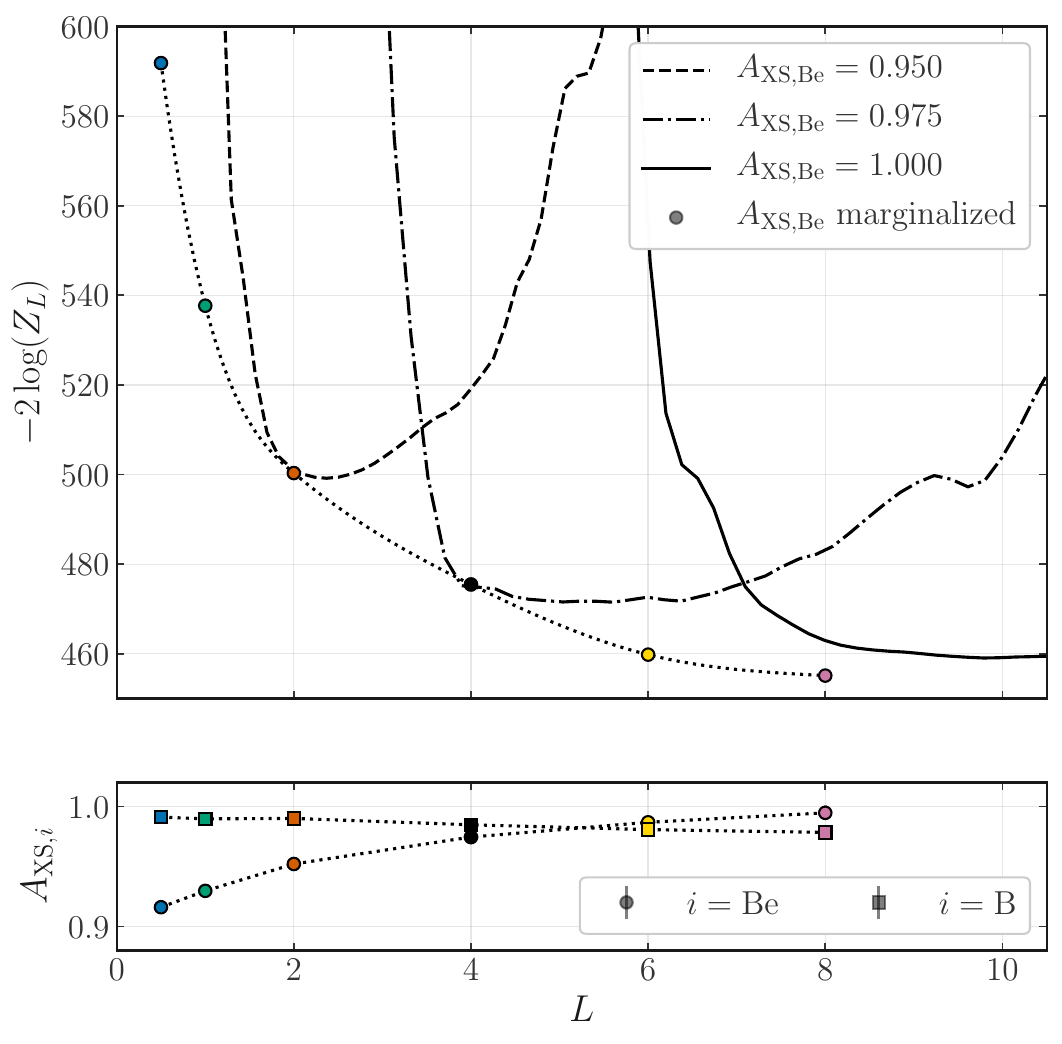}
    \caption{Evidence as function of $L$. The dots combine the results of the six fits of the {\tt Conv $v_{0,c}$} model to construct the posterior. Additionally, we show the evidences of $L$ for different fixed values of the Be cross section renormalization (this is rescaled from the posteriors, see text for details). In the bottom panel, we display the renormalization factors for Be and B production cross sections within the {\tt Conv $v_{0,c}$} model for different $L$, i.e. in correspondence of the dots in the upper panel. }\label{fig:chiBeB}
\end{figure}

In Fig.~\ref{fig:chiBeB} we show the interplay between the value of $L$ and the normalization of the beryllium cross section. We do this exercise for the {\tt Conv $v_{0,c}$} model. In the upper panel the points show the evidences  obtained from CR fits with $L$ as a fixed parameter. Instead, for the cases with fixed Be cross section normalization we allow $L$ as a free parameter. 
The connection between the evidence with free and fixed $L$ can be derived as follows.
Let us denote with $\bm\theta$ all fit parameters except $L$. If $L$ is a fixed parameter the evidence is given by:
\begin{eqnarray}
Z_L =  \int d \bm\theta \; p(D| \bm\theta,L) p( \bm\theta) \, . 
\end{eqnarray}
On the other hand, if $L$ is a free parameter in the fit, the posterior probability for $L$ is defined by
\begin{eqnarray}
p(L|D) &=& \frac{\int d \bm\theta \; p(D| \bm\theta,L) p( \bm\theta) p(L)} 
              {\int d \bm\theta dL \; p(D| \bm\theta,L) p( \bm\theta)p(L)}  \\ \nonumber
       &=& \frac{p(L)}{Z} \int d \bm\theta \; p(D| \bm\theta,L) p( \bm\theta) \, , 
\end{eqnarray}
assuming that the prior of $L$ factorizes (i.e. is uncorrelated) from $\bm\theta$.
It is thus possible to extract the equivalent of the evidence with fixed $L$:
\begin{equation}
\label{eq:evidence_fixed_L}
Z_L =  \frac{p(L|D) Z}{p(L)}   \, ,
\end{equation}
where
$Z$ is the evidence of the fits with free $L$. 

Among the tested cases for $L$ the best propagation model is the one for  $L=8$ kpc. In fact, we can see from the top panel of Fig.~\ref{fig:chiBeB} that the smaller is $L$ and the worse is the fit. 
We expect however that the Bayesian evidence $Z_L$ has a plateau for $L>8$ kpc.
Taking the statistical results for $L$ at face value, our findings can be used to put a frequentist lower limit for $L$ which is at the level of $4$ kpc at $5\sigma$ C.L. The ratio of the Bayesian evidence between the case with $L=4$ kpc and 8 kpc is about $2.6 \times 10^4$, similarly to the result obtained in the frequentist statistical framework on the lower limit for $L$.
This result is qualitatively compatible with the one shown in Ref.~\cite{Korsmeier:2021brc}.

The results for $L$ are affected by the uncertainties on the nuclear cross sections, in particular the ones for the beryllium production. 
The peculiarity here is the $\beta$-decay of $^{10}$Be to $^{10}$B in a $\tau_{1/2} = 1.37$ Myr, which alters both the Be and the B fluxes 
\cite{Donato:2001eq,Putze:2010zn,Maurin:2022gfm, Evoli:2019iih}. 
Given their short lifetime, the radioactive clocks such as $^{10}$Be  can be used to set bounds of the thickness of the diffusive halo \cite{Donato:2001eq}. 
The impact of $^{10}$Be is maximal in the 
$^{10}$Be/$^{9}$Be ratio, but can be sizeable also in the Be/B ratio. 

In  the bottom panel of Fig.~\ref{fig:chiBeB} we report the best-fit value for the parameter $A_{\rm{XS}}\rightarrow$ Be and $A_{\rm{XS}}\rightarrow$ B that we obtain when we perform the fit fixing the value of $L$ to different values. We remind that the parameters $A_{\rm{XS}}\rightarrow$ Be and $A_{\rm{XS}}\rightarrow$ B remormalize the nuclear cross sections implemented in \textsc{Galprop} for the Be and B production. We can see that $A_{\rm{XS}}\rightarrow$ Be takes values of the order of 0.95 when $L=0.5$ kpc and increases with $L$ reaching a plateau at 1 for $L>6$ kpc.
This is expected because the beryllium, and in particular $^{10}$Be, is the only isotope with a decay time comparable to the size of the diffusive halo.
Therefore, the exact value of $A_{\rm{XS}}\rightarrow$ Be can affect the best-fit value of $L$ in our results.

Due to the ignorance of the cross sections for the production of Be, B and Li, we marginalize over the cross section parameters by assuming they are nuisance parameters. 
In particular, values of these renormalizations of the order of $10\%$, which are reasonable given the current collider data, bring very different best-fit values of the diffusive halo.
In order to demonstrate this, we perform a fit to the data by fixing the cross sections for the production of B, Be and Li, and leaving free the value of $L$. We work with the model {\tt Conv $v_{0,c}$} and we fix $A_{\rm{XS}}\rightarrow$ B, $A_{\rm{XS}}\rightarrow$ Li, and $\delta_{\rm{XS}}\rightarrow$ B, Be, Li to the best-fit values reported in Tab.~\ref{tab:bestfit_L}.
For $A_{\rm{XS}}\rightarrow$ Be we test three different values of 0.95, 0.975 and 1.00.
We find that for this three possibilities the best fit value of $L$ are: $2.4^{+0.2}_{-0.3}$ kpc, $5.4^{+1.0}_{-0.7}$ and $10.8^{+1.3}_{-1.9}$ kpc, respectively.
The Bayesian log-evidences we obtain for each of the three tested cases, are $-247.9$, $-232.7$ and $-225.6$, respectively.

The Bayes factors between them show that the models with large $L$ are statistically favoured.
The propagation parameters are in good agreement with the values obtained for fixed $L$ at 2, 4 and 8 kpc.
Alternatively, in a frequentist interpretation the $\chi^2$ obtained in the three cases are 379, 364 and 339 leading to a similar conclusion.

We note however that a purely statistical interpretation of the $L$ dependence might not cover the whole story. The CR propagation model is phenomenological and not only derived from first principle. Therefore, some level of discrepancy between model and data is expected. This might lead to some bias which is then compensated by the cross section nuisance parameters. Thus a robust conclusion will rely on a better determination of the cross section. For example, if the Be cross sections turn out to be five percent smaller than the default assumed in this paper, $L$ will be constrained to smaller values around 2 kpc (see Fig.~\ref{fig:chiBeB}). In terms of the absolute $\chi^2$ also $L\in[2,4]$~kpc provides a good fit to the data.
So, all in all we find a statistical preference for large values of $L$ while noting that because of systematic effects also smaller values around 2 kpc should not be completely discarded.

\section{Discussion}
In this section we discuss our results in the context of recent literature on  secondary CR positrons, and we outline their broader implications. We also assess possible further uncertainties on our predictions. 
As a general caveat, we note that a precise comparison of our results with previous works is challenged by different treatment of many crucial ingredients, such as the propagation models and the production cross sections. 

A first comparison can be made with what obtained with \textsc{Galprop}  in \cite{Ma_2023}, and specifically with their model named B', which includes reacceleration and high rigidity break in the diffusion coefficient with $L=3.61$~kpc. Our results within {\tt Reacc Inj$_{30}$} model are lower by a factor of about 1.5 at 10~GeV and of about a factor of two at few GeV. As for the {\tt Conv $v_{0,c}$} model,  we obtain similar results at tens of GeV. At lower energies, their model ,including a reacceleration velocity of about 20~km/s, drives higher fluxes of  secondary IS positrons, larger by a factor up to 1.5  at 2~GeV, and indeed overshooting the AMS-02 data.  

Predictions obtained with the semi analytical propagation models SLIM, BIG and QUAINT  as defined in \cite{Weinrich:2020ftb} compare to our results as follows.
The case {\tt Conv $v_{0,c}$} with $L=4$~kpc is a factor of two larger at about 5~GeV with respect to their BIG-MED, which has zero reacceleration and a best-fit convection velocity around zero. Similar differences are found with respect to the SLIM model. When comparing the QUAINT model results with our {\tt Reacc Inj$_{30}$},
which both include significant reacceleration velocities, we consistently find lower positron fluxes. We note that these semi-analytical propagation models assume different shapes for the diffusion coefficients as a function of rigidity, and for the source terms, as well as of course the production cross sections, which can be the reasons of the discrepancies.

Further predictions for the secondary positrons at Earth have been obtained in \cite{Evoli_2021} with a semi-analytic model, using primary CRs fluxes from \cite{Boschini_2017}. Their results are very similar to our predictions within the  {\tt Reacc Inj$_{30}$} propagation model.  

A general consequence of the results illustrated in Sec.~\ref{sec:results} is that the predicted $e^+$ secondary contribution is not able to account for the AMS-02 data not even  around a few GeV. This is a theoretically challenging result, since the secondary contribution is typically assumed to explain the data up to 10 GeV (see, e.g., \cite{Hooper:2008kg,Serpico:2011wg,Tomassetti:2015mha,Delahaye:2008ua,Delahaye_2009,DiMauro:2014iia}).

We have demonstrated that the uncertainties related to the leptonic production cross section are now much smaller than the gap between the predicted secondaries and the positron flux data.
In fact, cross section uncertainties were considered to introduce an uncertainty of the order of 20-30\% \cite{Delahaye:2008ua,Evoli:2017vim,Evoli_2021}, which could have partially explained the mismatch at low energies.
Our results indicate that, within the propagation model explored here, an excess of positrons is present at energies larger than a few GeV, where the secondary flux starts to be less than $50\%$ than the data. 
While this is consistent with a number of previous works \cite{Boudaud:2016jvj,Weinrich:2020cmw,Fornieri_2020}, our results prove that for fixed values of $L \sim$ 4 kpc, positron cross sections uncertainties are too small to explain the mismatch at low energies.
However, we should notice that a larger secondary production is still not firmly excluded for smaller values of $L$, even if they correspond to worse fits to current nuclei CR data.  From a study of the nuclear fragmentation cross section, we can conclude that measurements for the nuclear cross sections involving the production of beryllium and its isotopes 
are needed with a precision below $5\%$ in order to estimate the size of the diffusive halo with a precision better than $50\%$. 

Further  uncertainties  may derive from leptonic energy losses, which above few GeV are dominated by inverse Compton and synchrotron emission. Updated estimates of the ISRF model 
in the solar neighborhood \cite{Vernetto:2016alq}, which well agree with the default \textsc{Galprop} model, reduce significantly the uncertainties in the ISRF provided by star and dust, as compared e.g. with the uncertainties parametrized in the M1-M3 models of \cite{Delahaye_2009}. 
In addition, we have verified that accounting for the 3D structure of the ISRF as recently modeled within \textsc{Galprop} \cite{Porter:2017vaa} by using the two benchmarks named F98 and R12, provides consistent results. The reason is due to the fact that the local photon densities are very constrained. 
Finally, a consistent estimate of the uncertainties in the synchrotron losses coming from the Galactic magnetic field model and its local value should proceed through a combined fit of the CR propagation models and of multi-wavelength data, such as radio, microwave and gamma-ray emissions \cite{Orlando:2019vmq}, which definitely would deserve a dedicated work. 

The ISM target gas density is another crucial ingredient for the computation of the secondary positrons in Eq.~\eqref{eqn:energy_spectrum}. The impact of updated models for the 3D ISM structure on CR was recently studied in Ref.~\cite{Johannesson:2018bit}, finding variations up to a factor of two for the column density of the local gas. 
In the analysis of CR nuclei data, we expect that the ISM density is 
effectively degenerate with the value of the diffusion coefficient. 
As a confirmation of this hint, we have verified that varying the ISM gas model among the ones available within \textsc{Galprop} \cite{Porter:2021tlr} in 2D and 3D, secondary CRs such as positrons are affected in the same way, and the ratio of secondary positrons to Boron remains constant to a good approximation.
These results suggest that the impact on the $e^\pm$ flux by varying the ISM as well as changing from  a 2D to a 3D modeling would be very moderate.

\section{Conclusions}
\label{sec:conclusions}
One of the strongest evidences for the presence of antimatter particles in our Galaxy is in the data of CR $e^+$, which reached a high precision on a wide energy range spanning from GeV to TeV \cite{PhysRevLett.122.041102}. An unavoidable contribution to $e^+$ in the Galaxy is due to the inelastic collisions of nuclei CRs - mainly $p$ and He - on the ISM atoms. This secondary source strongly depends on the hadronic cross sections at the basis of the processes.  
The knowledge of the secondary $e^+$ component in CRs is crucial to the understanding of this antimatter channel. 
A better determination of the secondary component and its uncertainties also implies
a more precise estimation of the room left by the data to any additional component. 
This gap  can in principle be ascribed to primary sources, such as pulsars or particle dark matter annihilation. 

In this paper we have provided a new prediction for the secondary $e^+$ flux in the Galaxy. We implement new $e^+$ production cross sections for $pp$ and $p$-nuclei collisions that became available recently \cite{Orusa:2022pvp}. 
In order to improve the Galactic propagation as well, we have performed new fits to CR nuclei data by computing the CR fluxes using \textsc{Galprop}, and have obtained new state-of-the-art propagation models. We test different propagation scenarios, characterized by specific choices on the diffusion coefficient, the convective wind, and reacceleration amount. We obtain very good fits to CR data, as quantified by the reduced $\tilde{\chi}$ smaller than 1 within each of the tested models. 
However, we find that propagation models with values of $L\lsim$ 2 kpc are disfavoured by CR data. 
We also study the consequences of nuisance parameters to 
allow some freedom in the fragmentation cross sections for the production of secondary CR nuclei. 

The results on the $e^+$ flux show that for all the propagation models selected by nuclei CR data, the 
$e^+$ flux never exceeds AMS-02 data. The excess of the data with respect to secondary $e^+$ production is significant from energies greater than few GeV. 
The $e^+$ flux at Earth depends in a significant amount on the size $L$ of the diffusive halo.
Models with $L\gsim$ 2 kpc can only explain the few AMS-02 data for positron energy $E< 1$ GeV.
We also assess the uncertainties on the $e^+$ flux due by propagation modeling and by production cross sections. The former are limited to 2-5\%, at fixed $L$ and depending on $E$, and are driven by the precision of AMS-02 nuclei data. A variation of $L$ from 8 to 2 (0.5) kpc 
implies a maximum rise of 50\% (250\%) in the propagated flux.
Uncertainties in the flux due to cross sections amount to  5-7\%, reflecting directly the results on the hadronic cross sections. This results reduces significantly this class of uncertainties with respect to the state of the art, and is a major finding of our work.

Contextually, we have computed the flux of secondary $e^-$ at Earth, following the same strategy as for $e^+$. As for $e^+$, the $e^-$ flux is determined with a high accuracy on the whole energy spectrum, thanks to the improvement in the determination of the hadronic cross sections, and the constraints on the propagation models. 
At $E< 1$~GeV, the  $e^-$ secondary fluxes is about 10\% of AMS-02 data, while for energies above few GeV the gap is about two orders of magnitude. This commonly known result is now reached with an unprecedented precision  well below 10\% on the whole energy spectrum, 
depending on 
the extension of the diffusive halo.

Summarizing, our results can be considered  new in a number of points:
i)
    The uncertainties on the positron flux attributed to inelastic hadronic cross sections are reduced to a few percent. We demonstrate that these uncertainties are nearly independent of the propagation setup.
ii) 
    We have calibrated the latest theoretical propagation model against a wide range of cosmic ray nuclei, obtaining updated parameters along with their corresponding uncertainties. Importantly, the size of the diffusion halo $L$ takes values between 2 and 8 kpc, while our analysis disfavors smaller values of $L$.
iii) 
    We have computed the positron flux utilizing the updated propagation models and quantified uncertainties. A special emphasize is placed on the size of the diffusion halo $L$ as it significantly impacts the prediction of the secondary positron flux. Our findings reveal that it is the current limiting factor for more accurate predictions of the positron flux. 
iv) 
    In analogy to positrons, we provide predictions for the flux of secondary electrons.

The results presented in this paper clearly indicate that 
a further better determination of $e^+$ flux - not necessarily due to secondary origin - is only possible after a more precise determination of the size of the region in which CRs are confined. An improvement in this direction could come, {\it i.e.} from precise data of radioactive isotopes such as the $^{10}$Be/$^{9}$Be ratio on a wide range of energies extending preferably above 20 GeV/n. 
CR positron measurements by the planned missions such as AMS-100 \cite{Schael:2019lvx} and Aladino \cite{Battiston:2021org} would permit to explore the secondary positron emission up to $\sim5$~TeV with percent statistical uncertainties.
An increased statistics in the measurement of positrons in the multi-TeV range could also help to break the degeneracy between the model's propagation parameters.

\begin{acknowledgments}
M.D.M., F.D. and L.O. acknowledge the support of the Research grant {\sc TAsP} (Theoretical Astroparticle Physics) funded by Istituto Nazionale di Fisica Nucleare. 
M.K. is supported by the Swedish Research Council under contracts 2019-05135 and 2022-04283 and the European Research Council under grant 742104.
S.M. acknowledges the European Union's Horizon Europe research and innovation programme for support under the Marie Sklodowska-Curie Action HE MSCA PF–2021,  grant agreement No.10106280, project \textit{VerSi}. 

The majority of the computation have been performed at the SLAC National Accelerator Laboratory Batch Farm, while some computation have been performed using the Swedish National Infrastructure for Computing (SNIC) under project Nos. 2021/3-42, 2021/6-326 and 2021-1-24 partially funded by the Swedish Research Council through grant no. 2018-05973.

\end{acknowledgments}

\bibliographystyle{apsrev4-1}
\bibliography{paper}

\clearpage
\newpage

\appendix

\section{Extended results for cosmic-ray propagation}
\label{app:tables}

In this section we report an additional discussion about the results we obtain for the propagation parameters and the fit to the CR flux data.

\begin{figure*}[t]

    \includegraphics[width=\textwidth]{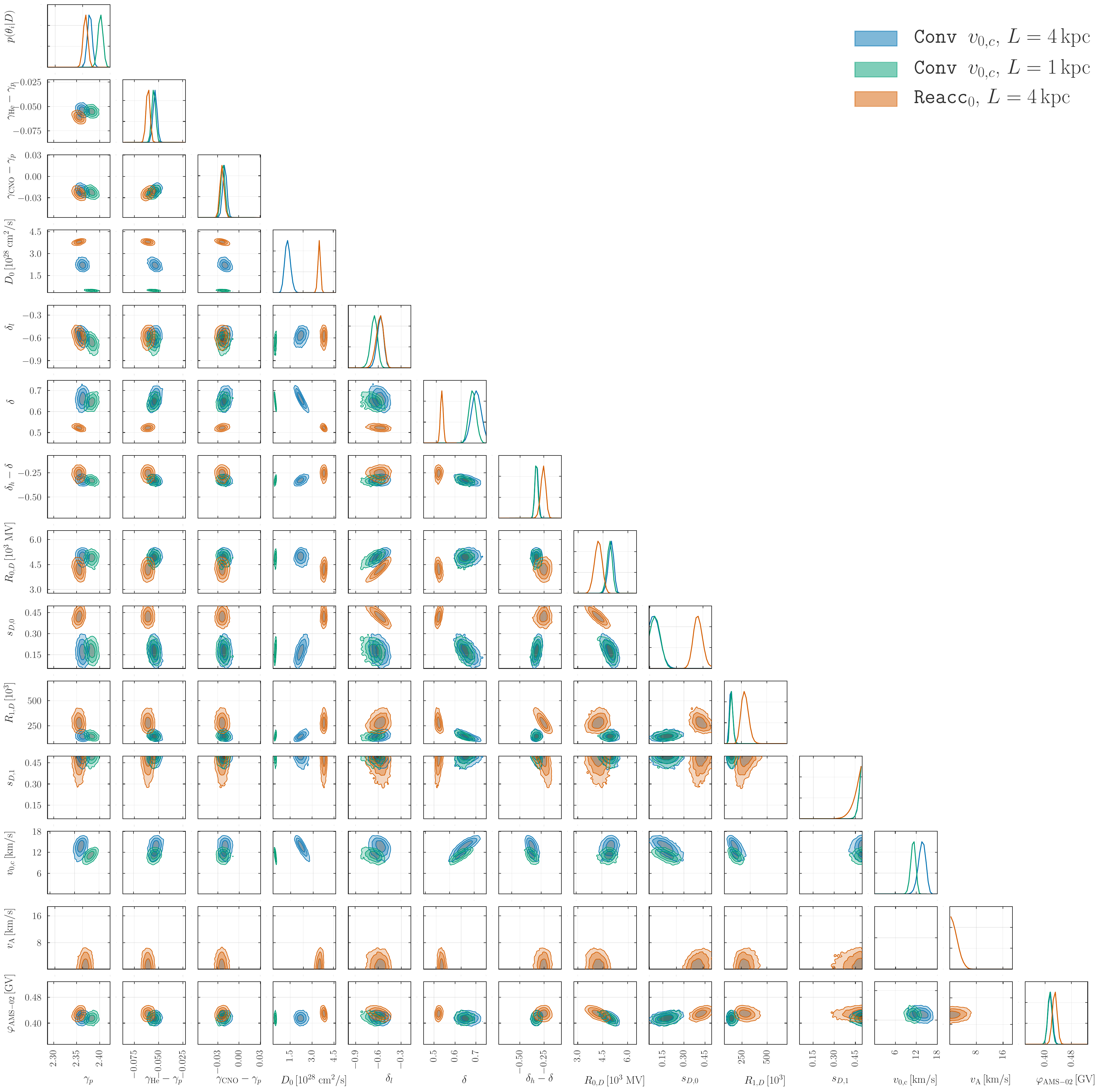}
    \caption{ Triangle plot for the propagation parameters obtained with the models \texttt{Conv $v_{0,c}$} with $L=1$ and 4 kpc and {\tt Reacc$_0$}. For each panel we show the $1$, 2 and 3$\sigma$ contours for each combination of two parameters, while the diagonal shows the posterior distribution for each individual parameter.}
    \label{fig:triangle1}
\end{figure*}

\begin{figure*}[t]
    \centering
    \includegraphics[width=\textwidth]{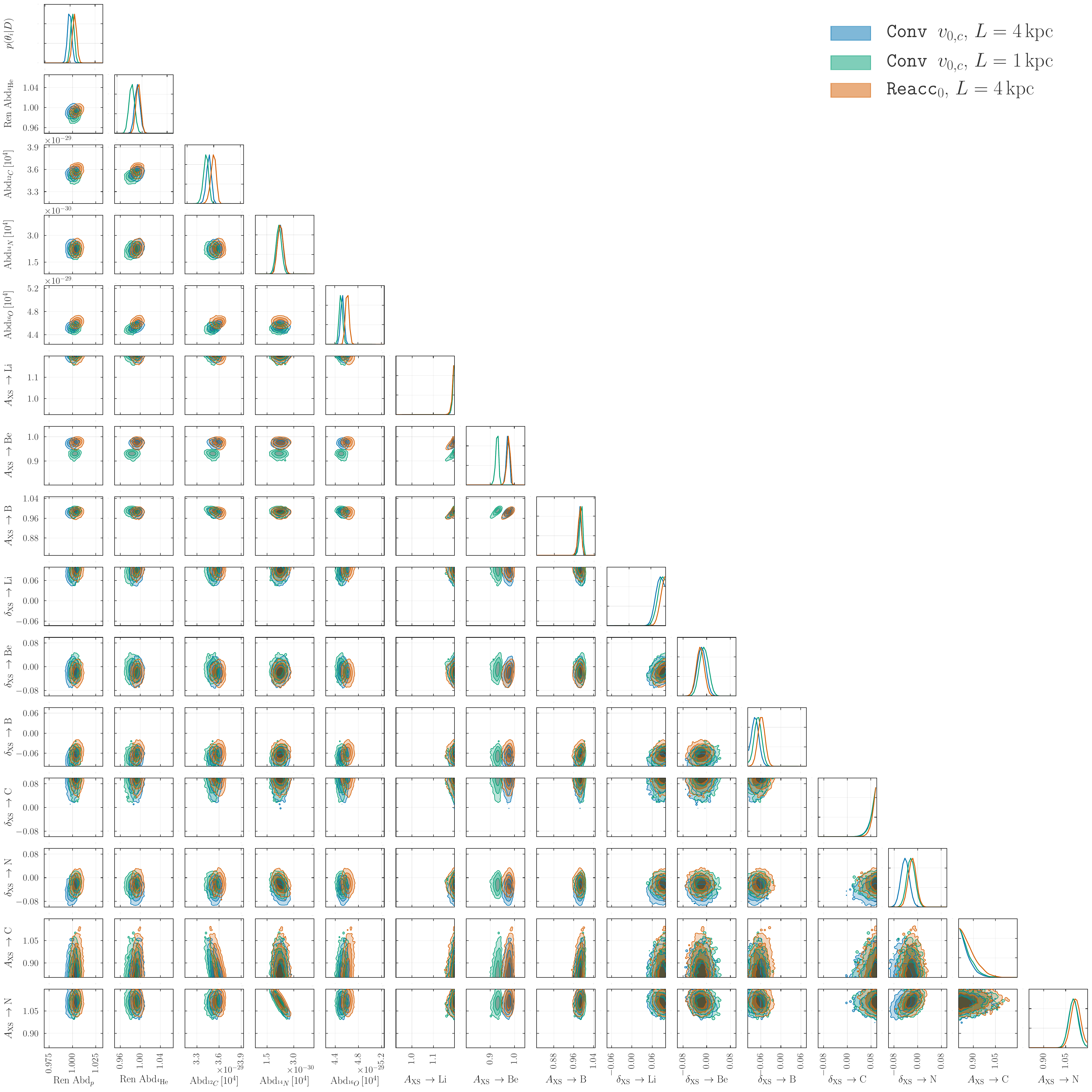}
    \caption{Same as Fig.~\ref{fig:triangle1} for the abundance of primary CRs and nuisance parameters of the cross sections.}
    \label{fig:triangle2}
\end{figure*}

In Tab.~\ref{tab:bestfit_L} we report the best-fit values obtained for the propagation parameters when we fix the model to the \texttt{Conv $v_{0,c}$} and we use  different values for $L$.
Additionally, in Tab.~\ref{tab:bestfit_model} we show the results we find when we use the models \texttt{Conv $v_{0,c}$}, \texttt{Conv $dv/dz$}, \texttt{Reacc$_0$}, \texttt{Reacc$_{10}$} and \texttt{Reacc$_{30}$ Inj}.
In Fig.~\ref{fig:triangle1}  (Fig.~\ref{fig:triangle2}) we show the triangle plots for the propagation 
(primary CRs abundance and nuisance parameters of the cross sections) parameters. We display the results obtained with the models \texttt{Conv $v_{0,c}$} with $L=1$ and 4 kpc, and {\tt Reacc$_0$}. For each panel we show the profiles and contours derived from the 1D and 2D marginalized posteriors. 

When we use \texttt{Conv $v_{0,c}$}, almost all the parameters found for different values of $L$ are compatible within the errors. The only exceptions are the value of $D_0$, which is proportional to $L$ (as explained in Sec.~\ref{sec:CRflux}), the convection velocity (see Fig.~\ref{fig:D0v0}) and the value of the normalization  cross sections for the beryllium production 
($A_{\rm{XS}}\rightarrow \rm{Be}$), see Fig.~\ref{fig:chiBeB}.
The slope we obtain for injections of protons $\gamma_p$ is about 2.36$-$2.39 while $\gamma{_{\rm He}}$ and $\gamma{_{\rm CNO}}$ are slightly softer of about $-0.05$ and $-0.02$, respectively.
The diffusion coefficient for the best-fit model increases below the first rigidity break at 5~GV ($\delta_l$ has a negative slope). The second slope $\delta$ is about $0.6-0.7$, while above the second break, located at around 155 GV, there is an hardening of about $\delta-\delta_h = 0.3$.
We find that there is a smooth transition of the diffusion coefficient between both breaks  with values of the smoothing of $0.15-0.20$ for the low-energy and 0.5 for the high-energy break.
The value we find for the slope $\delta$ is much larger than what typically is found in other references (see, e.g., \cite{Cuoco:2019kuu}) because indeed we include this smoothing also for the high-rigidity break. We also note that the best-fit for $s_{D,1}$ is at the edge of the prior (see Fig.~\ref{fig:triangle1}).
Talking about the nuisance parameters for the nuclear cross sections, the value of $A_{\rm{XS}}\rightarrow \rm{Li}$ is 1.20 and at the edge of the prior, as well as $\delta_{\rm{XS}}\rightarrow \rm{C}$ and $\delta_{\rm{XS}}\rightarrow \rm{Li}$ (see Fig.~\ref{fig:triangle2}).

When we use different propagation models, i.e.~the models with reacceleration, we find that leaving free  different slopes for p, He and CNO CRs, the fit improves significantly.
In particular, both the low and high-energy slopes of the spectra are slightly harder for He and CNO with respect to protons.
The best-fit parameters and the goodness of the fits found for the model \texttt{Conv $dv/dz$} are basically the same of the model \texttt{Conv $v_{0,c}$}.
The model labeled as {\tt Reacc$_0$} returns as best-fit value for $v_A$ about 0 km/s and the diffusion coefficient for the part above 10~GV is similar to the convective cases.

The triangle plots shown in Fig.~\ref{fig:triangle1} show the presence of correlations of a few parameters such as $D_0$ and $v_{0,c}$ (see also Fig.~\ref{fig:D0v0}). 
All the other parameters do not show strong correlations.

\begin{figure*}[t]

    \includegraphics[width=0.49\textwidth]{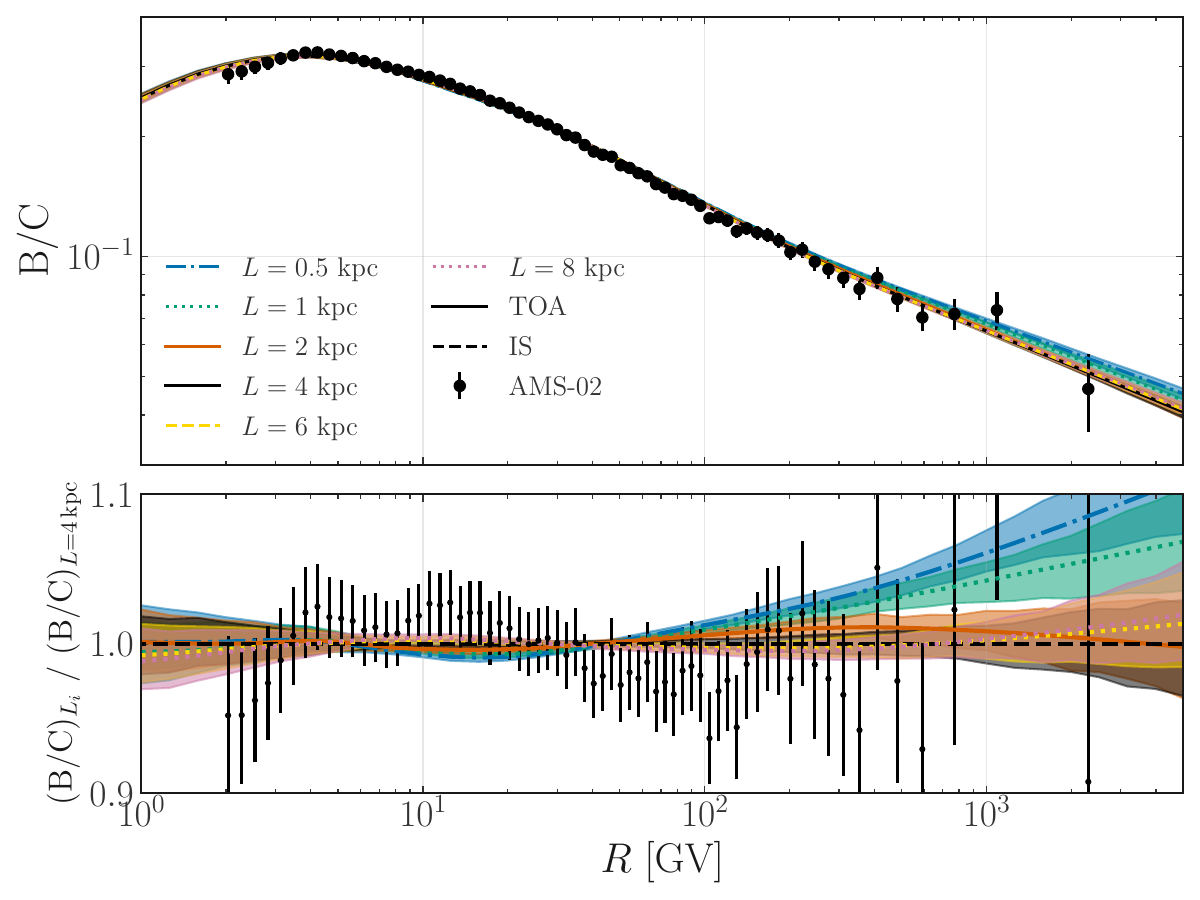}
    \includegraphics[width=0.49\textwidth]{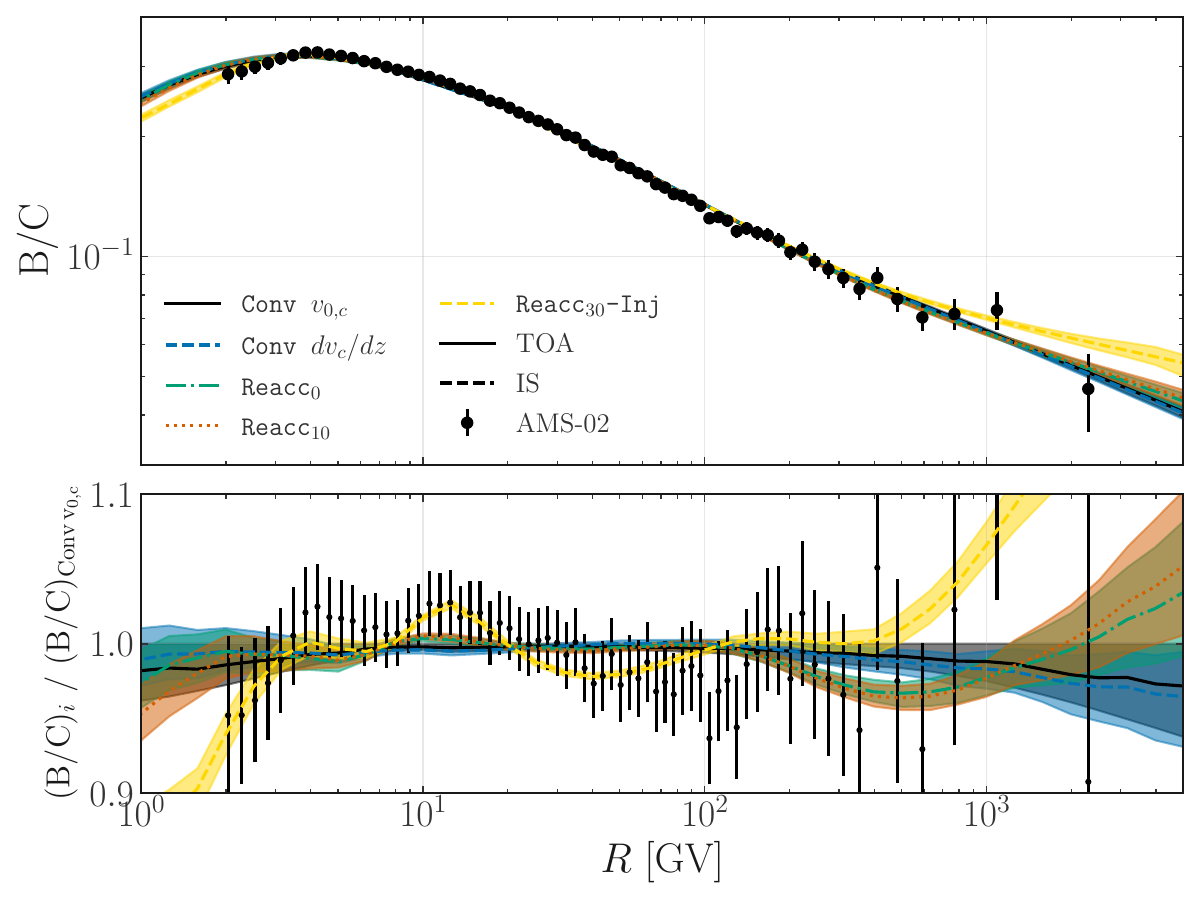}
    \includegraphics[width=0.49\textwidth]{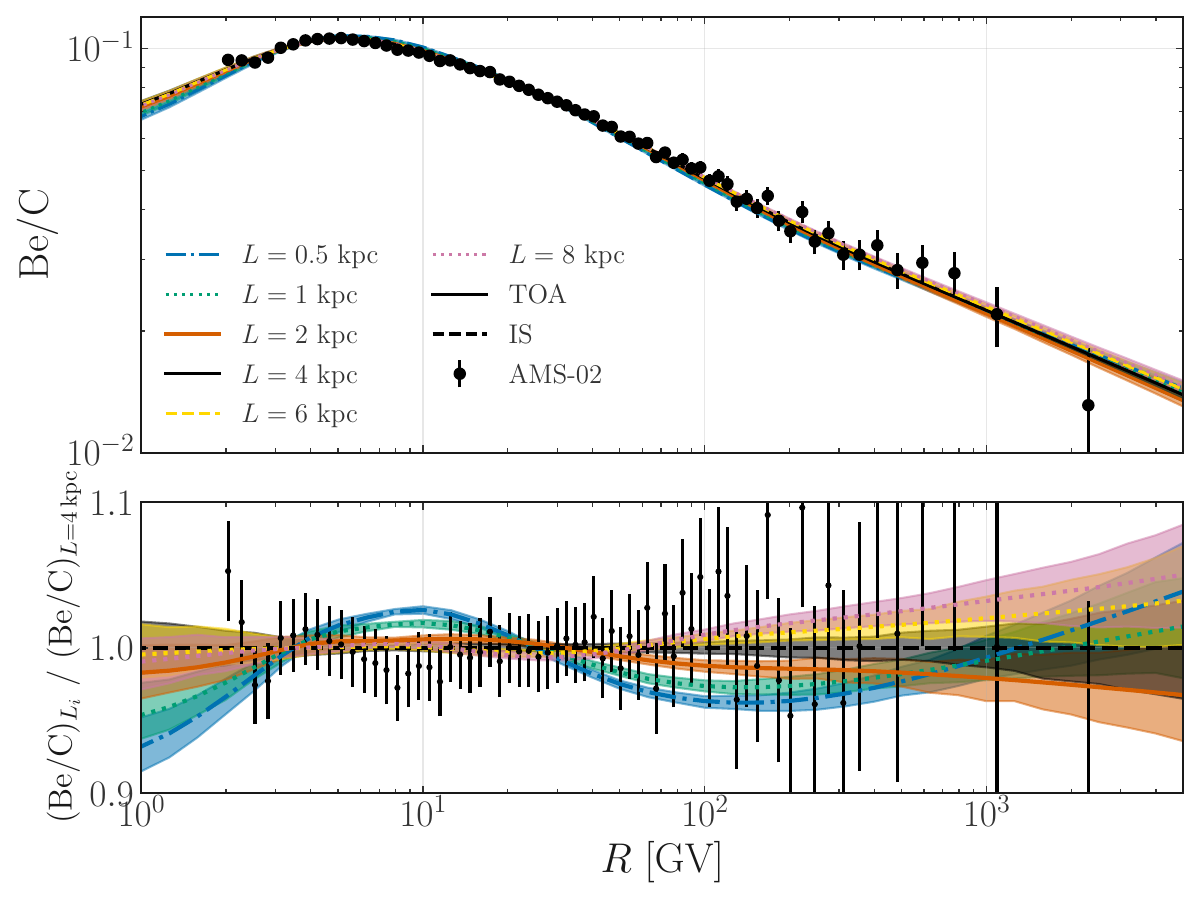}
    \includegraphics[width=0.49\textwidth]{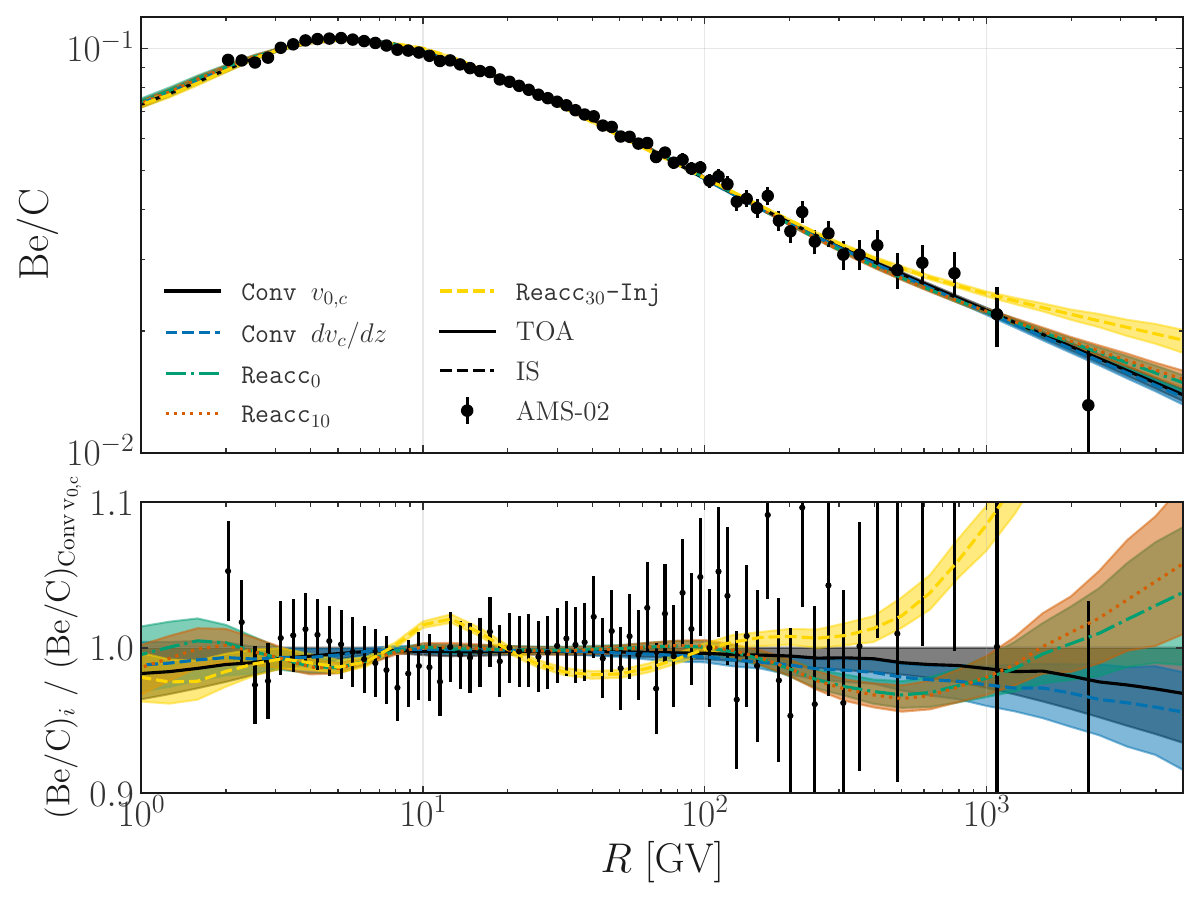}
    \caption{ Plot of the flux ratio between boron and carbon (top panel) and beryllium and carbon (bottom panel). In the left panels we show the results for the {\tt Conv $v_{0,c}$} model with different $L$, while in the right panel we report the other models tested in the paper. Below each figure we display the ratio between the different considered cases  and the result obtained for {\tt Conv $v_{0,c}$} with $L=4$ kpc. We also show the ratio between the data and the {\tt Conv $v_{0,c}$} model with $L=4$ kpc.}
    \label{fig:fluxsec}
\end{figure*}

In Fig.~\ref{fig:fluxsec} we show the ratio between the flux of secondary and primary CRs. In particular, we display the result for the ratio Be/C and B/C for the model {\tt Conv $v_{0,c}$} model for different values of $L$, and when we use the convection and reacceleration models.
All the tested models, with convection or reacceleration, provide a good fit to the secondary over primary ratios when we use $L=4$ kpc. In fact we see in the right panels that the differences between the tested models and the data are minor in the energy range of the data. The reduced $\chi^2$ found for all the models is below 1.

Instead, some differences are seen when we test different values of $L$. In particular, we can see in the plots that smaller values of $L$ are disfavored by the Be/C data for rigidities between a few GV up to tens of GV. The models with $L<2$~kpc struggle to fit the Be/C data and B/C data at the same time.  
Future AMS-02 data for the beryllium isotopes might help to put tight constraints for the size of the diffusive halo (see, e.g., \cite{Maurin:2022gfm}).

\begin{table*}[t]
    \centering
    \begin{tabular}{l rcl  ccccccccc }
\hline \hline
Parameter                        &
\multicolumn{3}{c}{Prior}        & 
$L=0.5$ kpc                      &
$L=1$ kpc                        &
$L=2$ kpc                        &
$L=4$ kpc                        &
$L=6$ kpc                        &
$L=8$ kpc                     
\\ \hline
$\gamma_p$                                                                 &                 2.2 &--& 2.5                  & $               {2.394}^{+0.005}_{-0.004}$ & $               {2.382}^{+0.005}_{-0.004}$ & $               {2.363}^{+0.004}_{-0.005}$ & $               {2.363}^{+0.004}_{-0.004}$ & $               {2.365}^{+0.004}_{-0.004}$ & $               {2.366}^{+0.004}_{-0.004}$ \\
$\gamma_{\rm{He}}-\gamma_p$                                                &                -0.1 &--& 0.1                  & $              {-0.056}^{+0.002}_{-0.002}$ & $              {-0.055}^{+0.002}_{-0.002}$ & $              {-0.053}^{+0.002}_{-0.002}$ & $              {-0.054}^{+0.002}_{-0.002}$ & $              {-0.054}^{+0.002}_{-0.002}$ & $              {-0.054}^{+0.002}_{-0.002}$ \\
$\gamma_{\rm{CNO}}-\gamma_p$                                               &                -0.1 &--& 0.1                  & $              {-0.024}^{+0.003}_{-0.003}$ & $              {-0.022}^{+0.003}_{-0.003}$ & $              {-0.020}^{+0.003}_{-0.002}$ & $              {-0.020}^{+0.002}_{-0.003}$ & $              {-0.020}^{+0.003}_{-0.003}$ & $              {-0.019}^{+0.003}_{-0.003}$ \\
$D_{0}\,\mathrm{[ 10^{28}\;cm^2/s]}$                                       &                 0.1 &--& 9.0                  & $               {0.225}^{+0.009}_{-0.012}$ & $                  {0.49}^{+0.02}_{-0.03}$ & $                  {1.01}^{+0.06}_{-0.06}$ & $                  {2.22}^{+0.12}_{-0.16}$ & $                  {3.20}^{+0.19}_{-0.20}$ & $                  {3.91}^{+0.22}_{-0.26}$ \\
$\delta_{l}$                                                               &                -1.0 &--& 0.0                  & $                 {-0.69}^{+0.05}_{-0.04}$ & $                 {-0.65}^{+0.05}_{-0.04}$ & $                 {-0.60}^{+0.04}_{-0.04}$ & $                 {-0.57}^{+0.04}_{-0.04}$ & $                 {-0.57}^{+0.04}_{-0.04}$ & $                 {-0.57}^{+0.04}_{-0.04}$ \\
$\delta_{}$                                                                &                 0.2 &--& 1.0                  & $                  {0.62}^{+0.01}_{-0.01}$ & $                  {0.64}^{+0.01}_{-0.02}$ & $                  {0.68}^{+0.02}_{-0.02}$ & $                  {0.66}^{+0.02}_{-0.02}$ & $                  {0.65}^{+0.01}_{-0.02}$ & $                  {0.65}^{+0.02}_{-0.02}$ \\
$\delta_h-\delta$                                                          &                -1.0 &--& 0.0                  & $                 {-0.32}^{+0.01}_{-0.02}$ & $                 {-0.33}^{+0.02}_{-0.01}$ & $                 {-0.34}^{+0.01}_{-0.02}$ & $                 {-0.33}^{+0.02}_{-0.02}$ & $                 {-0.32}^{+0.01}_{-0.02}$ & $                 {-0.32}^{+0.01}_{-0.02}$ \\
$R_{0,D}\,\mathrm{[ GV]}$                                          &                 1.0 &--& 10.0                 & $                  {4.84}^{+0.17}_{-0.17}$ & $                  {4.90}^{+0.15}_{-0.15}$ & $                  {4.97}^{+0.17}_{-0.11}$ & $                  {4.99}^{+0.17}_{-0.15}$ & $                  {4.99}^{+0.22}_{-0.10}$ & $                  {4.98}^{+0.17}_{-0.15}$ \\
$s_{D,0}$                                                                  &                 0.1 &--& 0.5                  & $                  {0.20}^{+0.03}_{-0.03}$ & $                  {0.18}^{+0.03}_{-0.03}$ & $                  {0.15}^{+0.04}_{-0.03}$ & $                  {0.17}^{+0.04}_{-0.04}$ & $                  {0.18}^{+0.03}_{-0.04}$ & $                  {0.18}^{+0.04}_{-0.03}$ \\
$R_{D,1}\;\mathrm{[ GV]}$                                         &                50.0 &--& 500.0                & $              {155.89}^{+13.68}_{-16.33}$ & $              {148.05}^{+12.92}_{-15.82}$ & $              {137.13}^{+15.27}_{-13.19}$ & $              {145.96}^{+13.91}_{-18.29}$ & $              {151.75}^{+13.70}_{-17.75}$ & $              {155.07}^{+14.82}_{-19.06}$ \\
$s_{D,1}$                                                                  &                 0.1 &--& 0.5                  & $               {0.485}^{+0.015}_{-0.003}$ & $               {0.486}^{+0.014}_{-0.002}$ & $               {0.488}^{+0.012}_{-0.002}$ & $               {0.486}^{+0.014}_{-0.002}$ & $               {0.486}^{+0.014}_{-0.003}$ & $               {0.485}^{+0.015}_{-0.003}$ \\
$v_{0,\mathrm{c}}\,\mathrm{[km/s]}$                                        &                 0.0 &--& 40.0                 & $                  {9.19}^{+0.62}_{-0.55}$ & $                 {11.13}^{+0.73}_{-0.67}$ & $                 {12.69}^{+0.81}_{-0.82}$ & $                 {13.67}^{+1.19}_{-0.92}$ & $                 {14.01}^{+1.13}_{-1.00}$ & $                 {14.07}^{+1.25}_{-1.04}$ \\
Ren Abd$_p$                                                                &                 0.9 &--& 1.1                  & $               {1.004}^{+0.002}_{-0.002}$ & $               {1.002}^{+0.002}_{-0.001}$ & $               {1.001}^{+0.002}_{-0.002}$ & $               {1.000}^{+0.002}_{-0.002}$ & $               {0.999}^{+0.002}_{-0.002}$ & $               {0.998}^{+0.002}_{-0.002}$  & * \\
Ren Abd$_{^4\rm{He}}$                                                      &                 0.9 &--& 1.1                  & $               {0.978}^{+0.004}_{-0.005}$ & $               {0.984}^{+0.004}_{-0.004}$ & $               {0.993}^{+0.004}_{-0.005}$ & $               {0.992}^{+0.004}_{-0.004}$ & $               {0.990}^{+0.004}_{-0.004}$ & $               {0.989}^{+0.004}_{-0.004}$  & * \\
Abd$_{^{12}C}\,[ 10^{4} ] $                                                &                 0.1 &--& 0.6                  & $               {0.349}^{+0.003}_{-0.003}$ & $               {0.351}^{+0.003}_{-0.003}$ & $               {0.356}^{+0.003}_{-0.002}$ & $               {0.355}^{+0.003}_{-0.002}$ & $               {0.354}^{+0.003}_{-0.002}$ & $               {0.354}^{+0.003}_{-0.002}$ \\
Abd$_{^{14}N}\,[ 10^{4} ] $                                                &                 0.0 &--& 0.1                  & $               {0.021}^{+0.002}_{-0.001}$ & $               {0.022}^{+0.001}_{-0.002}$ & $               {0.022}^{+0.002}_{-0.001}$ & $               {0.022}^{+0.002}_{-0.002}$ & $               {0.023}^{+0.001}_{-0.002}$ & $               {0.023}^{+0.002}_{-0.002}$ \\
Abd$_{^{16}O}\,[ 10^{4} ] $                                                &                 0.2 &--& 0.7                  & $               {0.449}^{+0.003}_{-0.003}$ & $               {0.450}^{+0.002}_{-0.003}$ & $               {0.454}^{+0.003}_{-0.003}$ & $               {0.453}^{+0.003}_{-0.003}$ & $               {0.452}^{+0.003}_{-0.003}$ & $               {0.452}^{+0.003}_{-0.002}$ \\
$A_\mathrm{XS}\,\rightarrow \mathrm{Li}$                                   &                 0.8 &--& 1.2                  & $              {1.197}^{+0.003}_{--0.001}$ & $              {1.197}^{+0.003}_{--0.001}$ & $              {1.197}^{+0.003}_{--0.001}$ & $              {1.198}^{+0.002}_{--0.001}$ & $              {1.198}^{+0.002}_{--0.001}$ & $              {1.197}^{+0.003}_{--0.001}$  & * \\
$A_\mathrm{XS}\,\rightarrow \mathrm{Be}$                                   &                 0.8 &--& 1.2                  & $               {0.916}^{+0.006}_{-0.004}$ & $               {0.930}^{+0.006}_{-0.005}$ & $               {0.952}^{+0.006}_{-0.004}$ & $               {0.975}^{+0.006}_{-0.004}$ & $               {0.987}^{+0.006}_{-0.004}$ & $               {0.995}^{+0.006}_{-0.004}$  & * \\
$A_\mathrm{XS}\,\rightarrow \mathrm{B}$                                    &                 0.8 &--& 1.2                  & $               {0.991}^{+0.005}_{-0.004}$ & $               {0.990}^{+0.006}_{-0.004}$ & $               {0.990}^{+0.005}_{-0.004}$ & $               {0.985}^{+0.006}_{-0.003}$ & $               {0.981}^{+0.006}_{-0.004}$ & $               {0.978}^{+0.006}_{-0.004}$  & * \\
$\delta_\mathrm{XS}\,\rightarrow \mathrm{Li}$                              &                -0.1 &--& 0.1                  & $               {0.085}^{+0.015}_{-0.004}$ & $               {0.086}^{+0.012}_{-0.006}$ & $               {0.084}^{+0.015}_{-0.006}$ & $               {0.082}^{+0.013}_{-0.008}$ & $               {0.084}^{+0.015}_{-0.005}$ & $               {0.085}^{+0.014}_{-0.005}$ \\
$\delta_\mathrm{XS}\,\rightarrow \mathrm{Be}$                              &                -0.1 &--& 0.1                  & $                 {-0.00}^{+0.02}_{-0.02}$ & $                 {-0.01}^{+0.02}_{-0.02}$ & $                 {-0.02}^{+0.01}_{-0.02}$ & $                 {-0.02}^{+0.02}_{-0.01}$ & $                 {-0.01}^{+0.01}_{-0.02}$ & $                 {-0.00}^{+0.02}_{-0.01}$ \\
$\delta_\mathrm{XS}\,\rightarrow \mathrm{B}$                               &                -0.1 &--& 0.1                  & $                 {-0.07}^{+0.01}_{-0.01}$ & $                 {-0.07}^{+0.01}_{-0.01}$ & $                 {-0.07}^{+0.01}_{-0.01}$ & $                 {-0.08}^{+0.01}_{-0.01}$ & $              {-0.074}^{+0.010}_{-0.012}$ & $                 {-0.07}^{+0.01}_{-0.01}$ \\
$\delta_\mathrm{XS}\,\rightarrow \mathrm{C}$                               &                -0.1 &--& 0.1                  & $               {0.088}^{+0.012}_{-0.002}$ & $               {0.086}^{+0.014}_{-0.003}$ & $               {0.086}^{+0.014}_{-0.002}$ & $               {0.086}^{+0.014}_{-0.002}$ & $               {0.086}^{+0.014}_{-0.003}$ & $               {0.085}^{+0.015}_{-0.003}$ \\
$\delta_\mathrm{XS}\,\rightarrow \mathrm{N}$                               &                -0.1 &--& 0.1                  & $                 {-0.02}^{+0.01}_{-0.01}$ & $                 {-0.02}^{+0.01}_{-0.01}$ & $                 {-0.03}^{+0.01}_{-0.01}$ & $                 {-0.04}^{+0.01}_{-0.01}$ & $                 {-0.04}^{+0.02}_{-0.01}$ & $                 {-0.04}^{+0.02}_{-0.01}$ \\
$\varphi_{\mathrm{AMS-02}}\,\mathrm{[GV]}$                                 &                 0.1 &--& 1.0                  & $               {0.416}^{+0.006}_{-0.007}$ & $               {0.415}^{+0.006}_{-0.007}$ & $               {0.414}^{+0.006}_{-0.006}$ & $               {0.416}^{+0.006}_{-0.007}$ & $               {0.416}^{+0.006}_{-0.006}$ & $               {0.417}^{+0.006}_{-0.006}$  & * \\
$\chi^2$ &  &  &  & 465 & 419  & 377  & 355  & 342 & 333  & \\
$\log(Z)$ &  & &   & -296  & -269  & -236  & -238  & -230  & -228  & \\
\hline
    \end{tabular}
    \caption{Summary of the best-fit parameters of the \texttt{Conv $v_{0,c}$} model for different sizes of the diffusion halo. The asterisk denotes that a parameter is sampled on-the-fly.
    The degeneracy between $L$ and $D_0$ is clearly visible. Actually, we take this also into account when choosing the prior for $D_0$. The value in the table states the maximal considered range. The priors for the individual fits are (in units of $10^{27}$ cm$^2$/s): [1,8], [2,10], [5,30], [8,50], [10,70], [15,90].}
    \label{tab:bestfit_L}
\end{table*}

\begin{table*}[t]
    \centering
    \begin{tabular}{l rcl  ccccccccc }
\hline \hline
Parameter                        &
\multicolumn{3}{c}{Prior}        & 
{\tt Conv $v_{0,c}$}             &
{\tt Conv $dv/dz$}               &
{\tt Reacc$_0$}                  &
{\tt Reacc$_{10}$}               &
{\tt Reacc$_{30}$-Inj}                                            
\\ 
\hline
$\gamma_{1,p}$                                                             &                 1.0 &--& 2.5                  & $ \gamma_{2,p} $       & $ \gamma_{2,p} $       & $ \gamma_{2,p} $       & $ \gamma_{2,p} $       & $                  {1.62}^{+0.03}_{-0.03}$ \\
$\gamma_{2,p}$                                                             &                 2.1 &--& 2.6                  & $               {2.363}^{+0.004}_{-0.004}$ & $               {2.354}^{+0.003}_{-0.005}$ & $               {2.355}^{+0.004}_{-0.004}$ & $               {2.357}^{+0.003}_{-0.004}$ & $               {2.440}^{+0.004}_{-0.005}$ \\
$\gamma_{1,\rm{He}}$                                                       &                 1.0 &--& 2.5                  & $ \gamma_{\rm{He}}-\gamma_p + \gamma_{2,p} $   & $ \gamma_{\rm{He}}-\gamma_p + \gamma_{2,p} $   & $ \gamma_{\rm{He}}-\gamma_p + \gamma_{2,p} $   & $ \gamma_{\rm{He}}-\gamma_p + \gamma_{2,p} $   & $                  {1.53}^{+0.04}_{-0.03}$ \\
$\gamma_{2,\rm{He}}$                                                       &                 2.1 &--& 2.6                  & $ \gamma_{\rm{He}}-\gamma_p + \gamma_{2,p} $   & $ \gamma_{\rm{He}}-\gamma_p + \gamma_{2,p} $   & $ \gamma_{\rm{He}}-\gamma_p + \gamma_{2,p} $   & $ \gamma_{\rm{He}}-\gamma_p + \gamma_{2,p} $   & $               {2.369}^{+0.004}_{-0.005}$ \\
$\gamma_{1,\rm{CNO}}$                                                      &                 1.0 &--& 2.5                  & $ \gamma_{\rm{CNO}}-\gamma_p + \gamma_{2,p} $   & $ \gamma_{\rm{CNO}}-\gamma_p + \gamma_{2,p} $   & $ \gamma_{\rm{CNO}}-\gamma_p + \gamma_{2,p} $   & $ \gamma_{\rm{CNO}}-\gamma_p + \gamma_{2,p} $   & $                  {1.77}^{+0.04}_{-0.04}$ \\
$\gamma_{2,\rm{CNO}}$                                                      &                 2.1 &--& 2.6                  & $ \gamma_{\rm{CNO}}-\gamma_p + \gamma_{2,p} $   & $ \gamma_{\rm{CNO}}-\gamma_p + \gamma_{2,p} $   & $ \gamma_{\rm{CNO}}-\gamma_p + \gamma_{2,p} $   & $ \gamma_{\rm{CNO}}-\gamma_p + \gamma_{2,p} $   & $               {2.377}^{+0.004}_{-0.006}$ \\
$R_{\rm{inj}}\;\mathrm{[ GV]}$                                    &                 0.5 &--& 10.0                 & --                                         & --                                         & --                                         & --                                         & $                  {5.68}^{+0.30}_{-0.36}$ \\
$s_{\rm{inj}}$                                                             &                 0.1 &--& 1.0                  & --                                         & --                                         & --                                         & --                                         & $                  {0.38}^{+0.02}_{-0.02}$ \\
$\gamma_{\rm{He}}-\gamma_p$                                                &                -0.2 &--& 0.1                  & $              {-0.054}^{+0.002}_{-0.002}$ & $              {-0.054}^{+0.002}_{-0.002}$ & $              {-0.061}^{+0.002}_{-0.002}$ & $              {-0.063}^{+0.002}_{-0.002}$ & --                                         \\
$\gamma_{\rm{CNO}}-\gamma_p$                                               &                -0.2 &--& 0.1                  & $              {-0.020}^{+0.002}_{-0.003}$ & $              {-0.021}^{+0.003}_{-0.003}$ & $              {-0.024}^{+0.003}_{-0.003}$ & $              {-0.025}^{+0.003}_{-0.003}$ & --                                         \\
$D_{0}\,\mathrm{[ 10^{28}\;cm^2/s]}$                                       &                 0.5 &--& 10.0                 & $                  {2.22}^{+0.12}_{-0.16}$ & $                  {2.21}^{+0.18}_{-0.19}$ & $                  {3.80}^{+0.06}_{-0.05}$ & $                  {3.97}^{+0.06}_{-0.04}$ & $                  {7.07}^{+0.25}_{-0.17}$ \\
$\delta_{l}$                                                               &                -2.0 &--& 0.5                  & $                 {-0.57}^{+0.04}_{-0.04}$ & $                 {-0.57}^{+0.05}_{-0.04}$ & $                 {-0.58}^{+0.05}_{-0.04}$ & $                 {-0.62}^{+0.04}_{-0.04}$ & $ \delta$              \\
$\delta_{}$                                                                &                 0.1 &--& 1.5                  & $                  {0.66}^{+0.02}_{-0.02}$ & $                  {0.66}^{+0.02}_{-0.03}$ & $               {0.522}^{+0.005}_{-0.006}$ & $               {0.509}^{+0.004}_{-0.005}$ & $               {0.390}^{+0.005}_{-0.004}$ \\
$\delta_h-\delta$                                                          &                -1.5 &--& 0.0                  & $                 {-0.33}^{+0.02}_{-0.02}$ & $                 {-0.33}^{+0.02}_{-0.02}$ & $                 {-0.26}^{+0.03}_{-0.02}$ & $                 {-0.25}^{+0.03}_{-0.03}$ & $                 {-0.22}^{+0.06}_{-0.03}$ \\
$R_{0,D}\,\mathrm{[ GV]}$                                          &                 0.5 &--& 10.0                 & $                  {4.99}^{+0.17}_{-0.15}$ & $                  {4.84}^{+0.18}_{-0.17}$ & $                  {4.22}^{+0.24}_{-0.22}$ & $                  {3.80}^{+0.12}_{-0.20}$ & --                                         \\
$s_{D,0}$                                                                  &                 0.1 &--& 0.5                  & $                  {0.17}^{+0.04}_{-0.04}$ & $                  {0.22}^{+0.03}_{-0.03}$ & $                  {0.42}^{+0.02}_{-0.03}$ & $               {0.478}^{+0.022}_{-0.006}$ & --                                         \\
$R_{D,1}\;\mathrm{[ GV]}$                                         &                50.0 &--& 800.0                & $              {145.96}^{+13.91}_{-18.29}$ & $              {156.24}^{+14.36}_{-19.32}$ & $              {281.77}^{+34.69}_{-38.27}$ & $              {318.29}^{+36.49}_{-48.36}$ & $             {415.79}^{+50.23}_{-104.50}$ \\
$s_{D,1}$                                                                  &                 0.1 &--& 0.5                  & $               {0.486}^{+0.014}_{-0.002}$ & $               {0.484}^{+0.016}_{-0.003}$ & $               {0.461}^{+0.039}_{-0.008}$ & $                  {0.45}^{+0.05}_{-0.01}$ & $                  {0.32}^{+0.10}_{-0.08}$ \\
$v_{0,\mathrm{c}}\,\mathrm{[km/s]}$                                        &                 0.0 &--& 40.0                 & $                 {13.67}^{+1.19}_{-0.92}$ & --                                         & --                                         & --                                         & --                                         \\
$dv/dz$ [km/s/kpc]                                                              &                 0.0 &--& 40.0                 & --                                         & $                 {11.71}^{+1.38}_{-1.39}$ & --                                         & --                                         & --                                         \\
$v_{\mathrm{A}}\,\mathrm{[km/s]}$                                          &                 0.0 &--& 100.0                & --                                         & --                                         & $                  {1.59}^{+0.41}_{-1.59}$ & --                                         & $                 {32.16}^{+1.56}_{-1.34}$ \\
Ren Abd$_p$                                                                &                 0.9 &--& 1.1                  & $               {1.000}^{+0.002}_{-0.002}$ & $               {0.998}^{+0.002}_{-0.002}$ & $               {1.005}^{+0.002}_{-0.002}$ & $               {1.005}^{+0.002}_{-0.002}$ & $               {1.030}^{+0.002}_{-0.002}$  & * \\
Ren Abd$_{^4\rm{He}}$                                                      &                 0.9 &--& 1.1                  & $               {0.992}^{+0.004}_{-0.004}$ & $               {1.000}^{+0.004}_{-0.004}$ & $               {0.994}^{+0.003}_{-0.004}$ & $               {0.997}^{+0.004}_{-0.003}$ & $               {0.946}^{+0.004}_{-0.005}$  & * \\
Abd$_{^{12}C}\,[ 10^{4} ] $                                                &                 0.1 &--& 0.8                  & $               {0.355}^{+0.003}_{-0.002}$ & $               {0.358}^{+0.003}_{-0.002}$ & $               {0.360}^{+0.003}_{-0.003}$ & $               {0.360}^{+0.004}_{-0.003}$ & $               {0.344}^{+0.005}_{-0.003}$ \\
Abd$_{^{14}N}\,[ 10^{4} ] $                                                &                 0.0 &--& 0.1                  & $               {0.022}^{+0.002}_{-0.002}$ & $               {0.022}^{+0.001}_{-0.002}$ & $               {0.023}^{+0.002}_{-0.002}$ & $               {0.022}^{+0.002}_{-0.002}$ & $               {0.025}^{+0.001}_{-0.002}$ \\
Abd$_{^{16}O}\,[ 10^{4} ] $                                                &                 0.1 &--& 0.8                  & $               {0.453}^{+0.003}_{-0.003}$ & $               {0.458}^{+0.002}_{-0.003}$ & $               {0.461}^{+0.003}_{-0.003}$ & $               {0.463}^{+0.003}_{-0.003}$ & $               {0.403}^{+0.002}_{-0.004}$ \\
$A_\mathrm{XS}\,\rightarrow \mathrm{Li}$                                   &                 0.8 &--& 1.2                  & $              {1.198}^{+0.002}_{--0.001}$ & $              {1.198}^{+0.002}_{--0.001}$ & $              {1.197}^{+0.003}_{--0.000}$ & $              {1.197}^{+0.003}_{--0.001}$ & $                  {1.96}^{+0.04}_{-0.01}$  & * \\
$A_\mathrm{XS}\,\rightarrow \mathrm{Be}$                                   &                 0.8 &--& 1.2                  & $               {0.975}^{+0.006}_{-0.004}$ & $               {0.971}^{+0.006}_{-0.004}$ & $               {0.978}^{+0.007}_{-0.004}$ & $               {0.980}^{+0.007}_{-0.004}$ & $                  {1.52}^{+0.04}_{-0.01}$  & * \\
$A_\mathrm{XS}\,\rightarrow \mathrm{B}$                                    &                 0.8 &--& 1.2                  & $               {0.985}^{+0.006}_{-0.003}$ & $               {0.986}^{+0.005}_{-0.004}$ & $               {0.985}^{+0.007}_{-0.004}$ & $               {0.986}^{+0.006}_{-0.004}$ & $                  {1.49}^{+0.04}_{-0.02}$  & * \\
$\delta_\mathrm{XS}\,\rightarrow \mathrm{Li}$                              &                -0.1 &--& 0.1                  & $               {0.082}^{+0.013}_{-0.008}$ & $               {0.080}^{+0.015}_{-0.007}$ & $               {0.090}^{+0.010}_{-0.003}$ & $               {0.090}^{+0.010}_{-0.003}$ & $               {0.292}^{+0.008}_{-0.002}$ \\
$\delta_\mathrm{XS}\,\rightarrow \mathrm{Be}$                              &                -0.1 &--& 0.1                  & $                 {-0.02}^{+0.02}_{-0.01}$ & $                 {-0.02}^{+0.01}_{-0.01}$ & $                 {-0.02}^{+0.01}_{-0.01}$ & $                 {-0.02}^{+0.01}_{-0.01}$ & $                  {0.24}^{+0.01}_{-0.01}$ \\
$\delta_\mathrm{XS}\,\rightarrow \mathrm{B}$                               &                -0.1 &--& 0.1                  & $                 {-0.08}^{+0.01}_{-0.01}$ & $              {-0.076}^{+0.010}_{-0.011}$ & $              {-0.059}^{+0.010}_{-0.011}$ & $              {-0.055}^{+0.010}_{-0.011}$ & $               {0.176}^{+0.009}_{-0.012}$ \\
$\delta_\mathrm{XS}\,\rightarrow \mathrm{C}$                               &                -0.1 &--& 0.1                  & $               {0.086}^{+0.014}_{-0.002}$ & $               {0.089}^{+0.011}_{-0.002}$ & $               {0.090}^{+0.010}_{-0.002}$ & $               {0.090}^{+0.010}_{-0.002}$ & $                  {0.21}^{+0.09}_{-0.03}$ \\
$\delta_\mathrm{XS}\,\rightarrow \mathrm{N}$                               &                -0.1 &--& 0.1                  & $                 {-0.04}^{+0.01}_{-0.01}$ & $                 {-0.03}^{+0.01}_{-0.01}$ & $                 {-0.02}^{+0.01}_{-0.01}$ & $                 {-0.01}^{+0.01}_{-0.01}$ & $                  {0.14}^{+0.02}_{-0.02}$ \\
$\varphi_{\mathrm{AMS-02}}\,\mathrm{[GV]}$                                 &                 0.1 &--& 1.0                  & $               {0.416}^{+0.006}_{-0.007}$ & $               {0.417}^{+0.005}_{-0.008}$ & $               {0.429}^{+0.007}_{-0.007}$ & $               {0.435}^{+0.006}_{-0.007}$ & $               {0.594}^{+0.008}_{-0.008}$  & * \\
$\chi^2$ &  &  &  &  355 &  399 & 437 & 465  & 264  &  \\
$\log(Z)$ &  & &    & -238  &  -263 & -278  & -290 &  -208  & \\
\hline
    \end{tabular}
    \caption{Same as Tab.~\ref{tab:bestfit_L} for the different models tested in the paper. The priors we use for the normalization and slope cross section parameters with the model {\tt Reacc$_{30}$ Inj} are $0.5-2.0$ and $-0.3,0.3$, i.e.~larger with respect to the ones used for the other models.}
    \label{tab:bestfit_model}
\end{table*}

\section{Grid tests}
In this section we expand the discussion on the choice of the grid for the numerical solution of the transport equation and show tests that we perform to find the optimal choice for the grid parameters.

\textsc{Galprop} solves the propagation equation numerically on a grid in $r$, $z$ and CR kinetic energy per nucleon ($E_{\rm kin}$) by updating the CR densities for discrete time steps. The properties of the  grids and the time step are defined as follows.
On the other hand, the grid in $E_{\rm kin}$ is logarithmic with a constant factor $f_{E_{\rm kin}}=E_{{\rm kin},{i+1}}/E_{{\rm kin},{i}}$. 
Finally, the time steps are defined in a more involved way.  \textsc{Galprop} solves the propagation equation starting with a large time step to quickly converge to an approximate solution and then logarithmically reduces the time step to converge to the accurate solution. Therefore, the following parameters can be defined: a starting time step and final time step, the number of repetitions at each time step and the time step factor (analogous to the $f_{E_{\rm kin}}$).
We note that also the minimal value for the $E_{\rm kin}$ grid is an important quantity. We use 1 MeV. A value larger than 10 MeV can have a significant impact also on the spectrum above 1 GeV because the CR density if forced to be 0 at the grid boundary. In particluar, this is important for models with reacceleration while for the {\tt Conv $v_{0,c}$} models the effect is smaller.

For the fits shown in the main text of the paper, we fix the grid by choosing the following values: $d z= 0.1$ kpc, $d r=1$ kpc, $f_{E_{\rm kin}}=1.1$, starting and ending time step of $10^9$ and $10^2$ years, time step factors of 0.5 and 20 repetitions.
These choices allow for a reasonably fast evaluation of \textsc{Galprop} while keeping the systematics at the level of a few percent.

\begin{figure*}
    \includegraphics[height=0.105\textwidth, trim={0     3cm 0 0},clip]{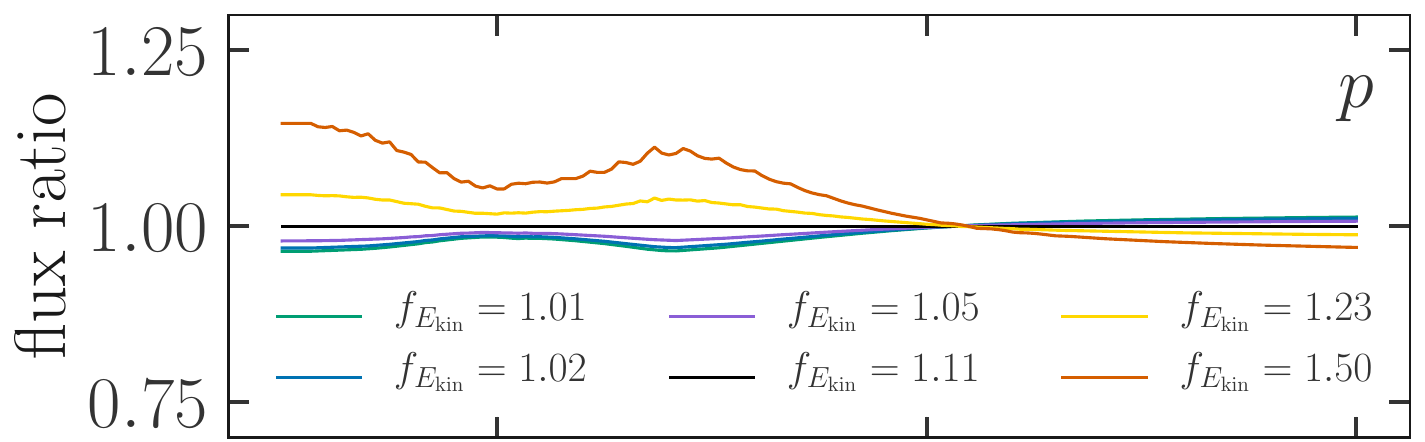}\hspace{0.01cm}
    \includegraphics[height=0.105\textwidth, trim={3.7cm 3cm 0 0},clip]{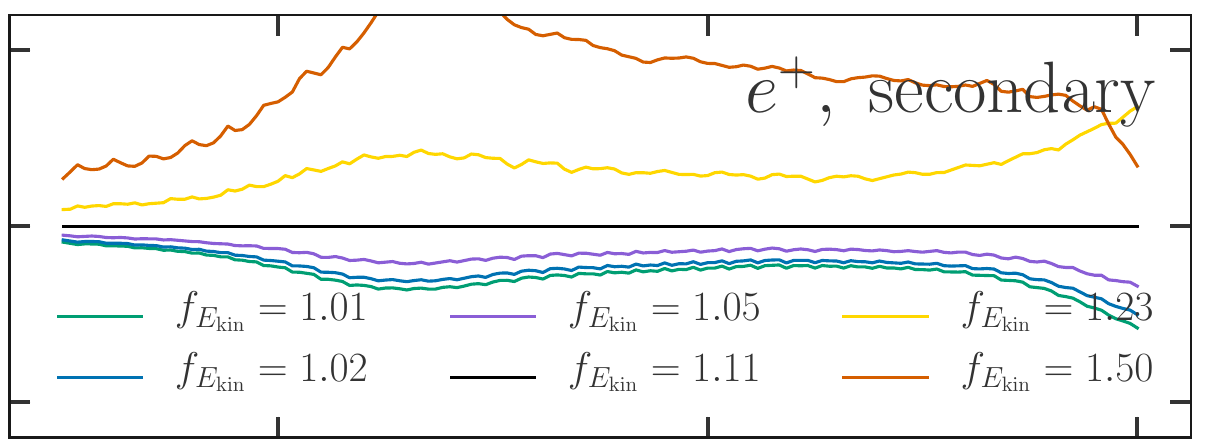}
    \includegraphics[height=0.105\textwidth, trim={3.7cm 3cm 0 0},clip]{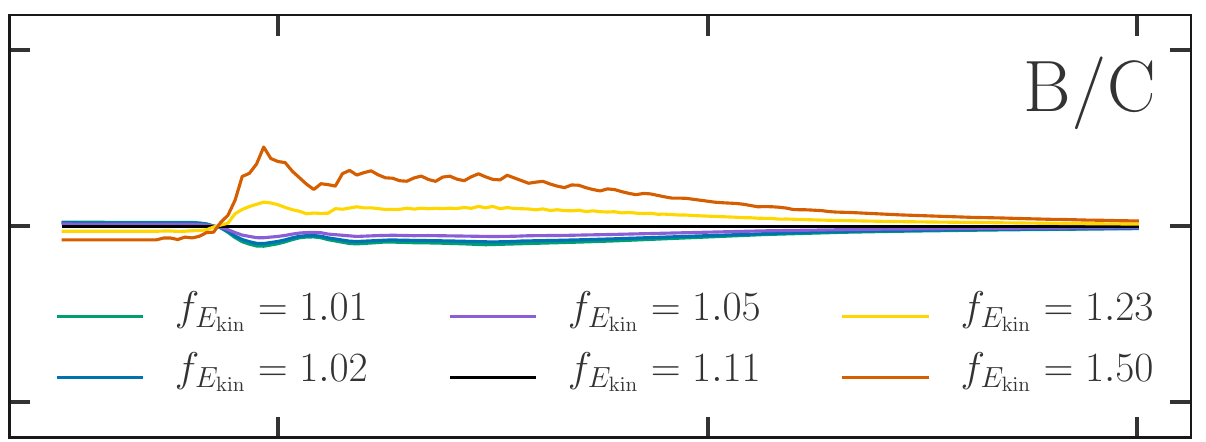}\vspace{0.13cm} \\
    \includegraphics[height=0.105\textwidth, trim={0     3cm 0 0},clip]{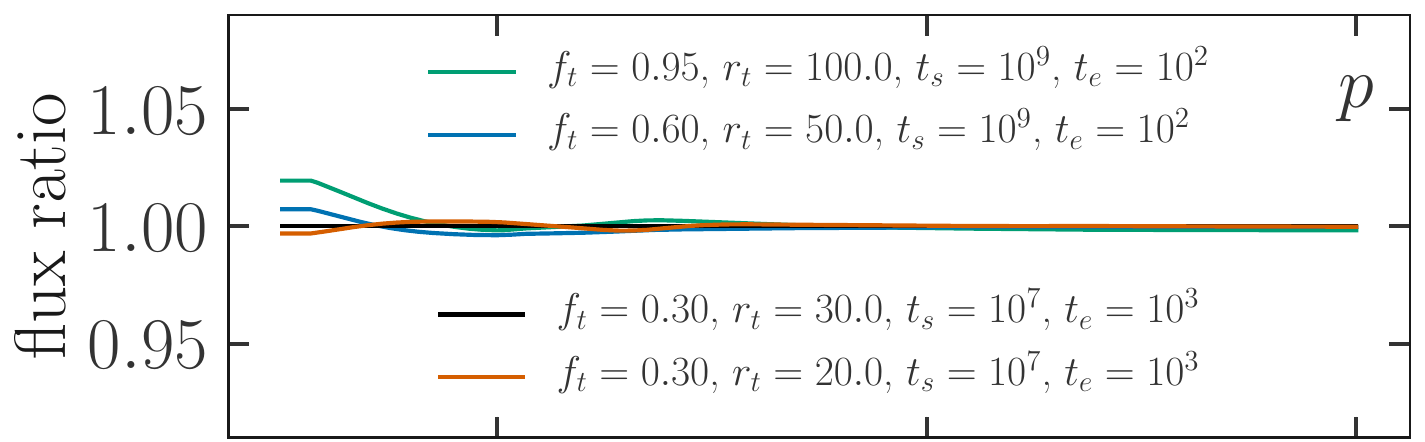}\hspace{0.01cm}
    \includegraphics[height=0.105\textwidth, trim={3.7cm 3cm 0 0},clip]{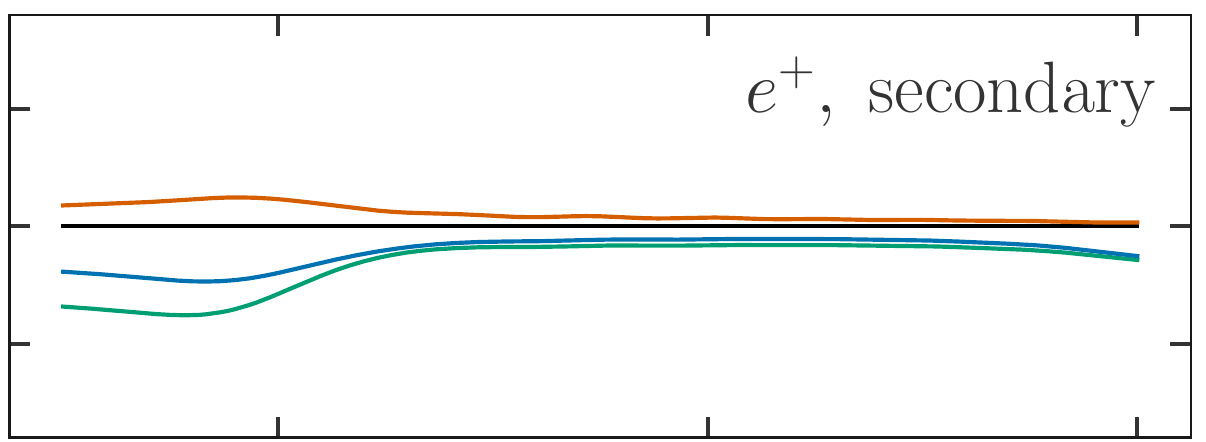}
    \includegraphics[height=0.105\textwidth, trim={3.7cm 3cm 0 0},clip]{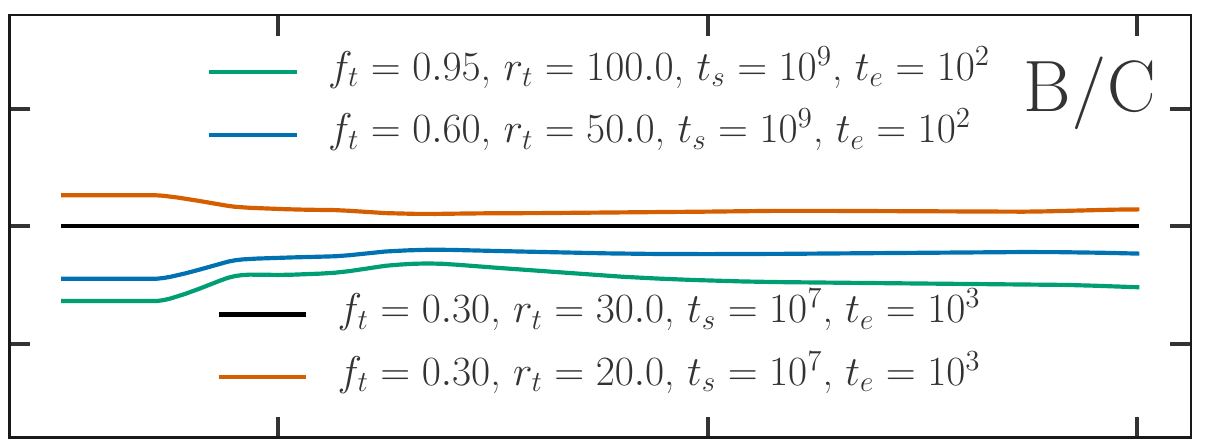}\vspace{0.13cm} \\
    \includegraphics[height=0.105\textwidth, trim={0     3cm 0 0},clip]{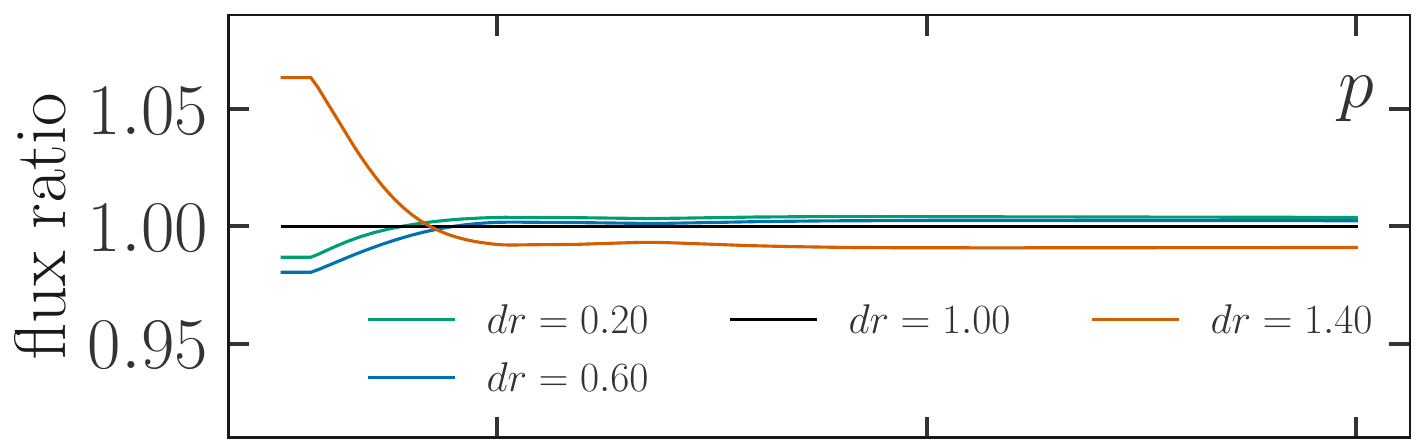}\hspace{0.01cm}
    \includegraphics[height=0.105\textwidth, trim={3.7cm 3cm 0 0},clip]{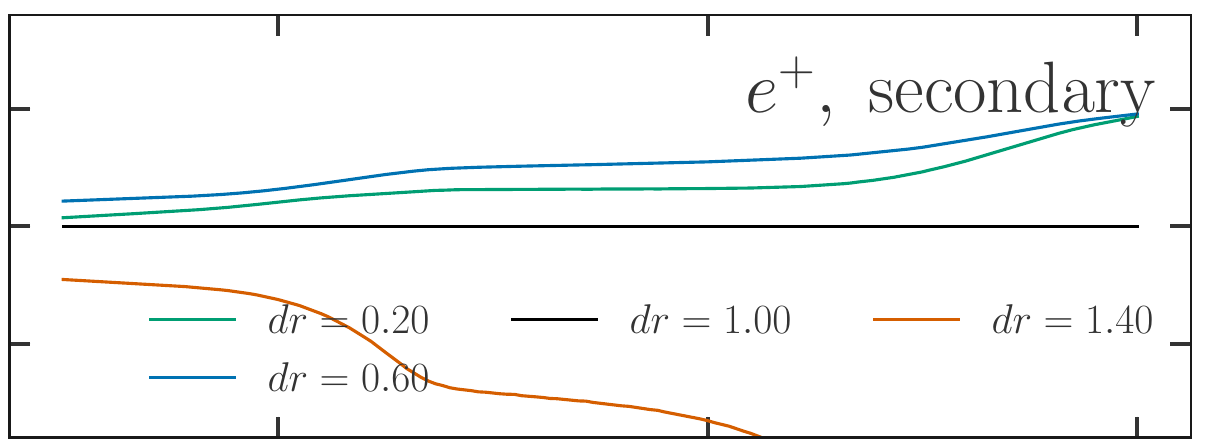}
    \includegraphics[height=0.105\textwidth, trim={3.7cm 3cm 0 0},clip]{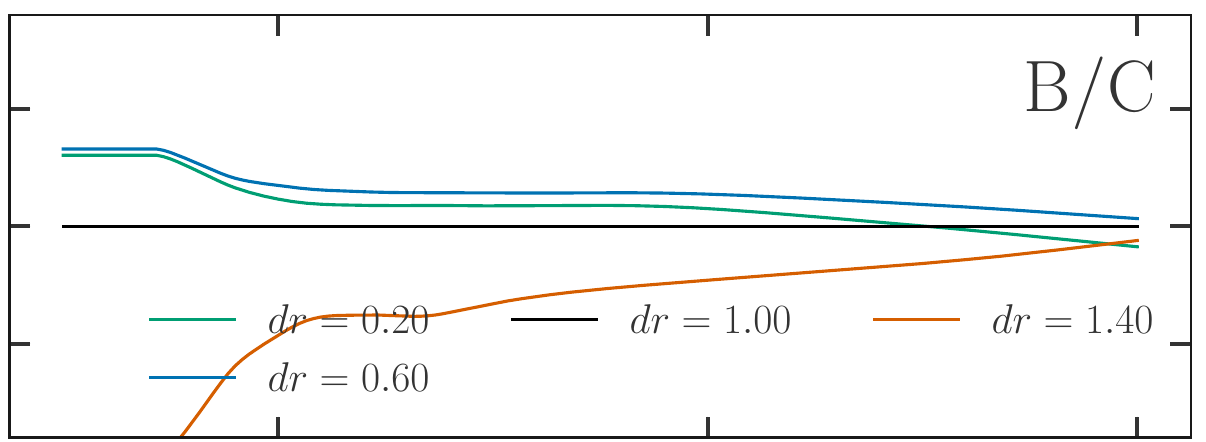}\vspace{0.13cm} \\
    \includegraphics[height=0.147\textwidth, trim={0     0 0 0},clip]{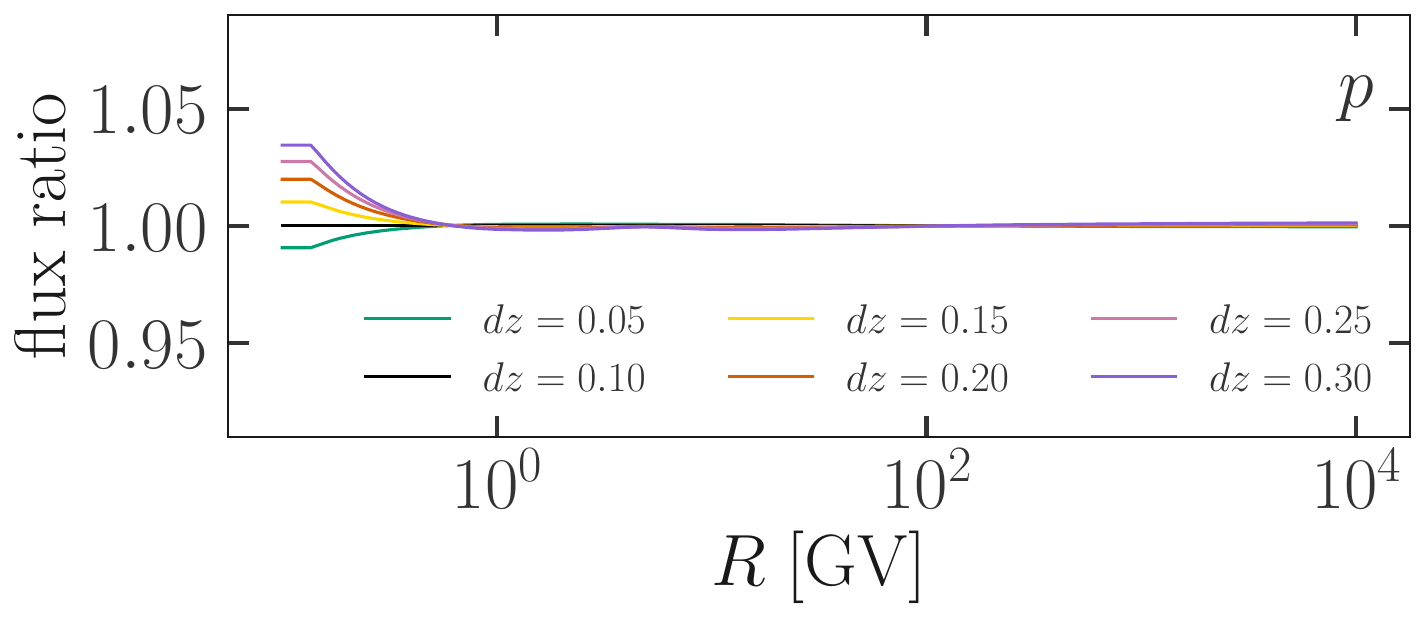}\hspace{0.01cm}
    \includegraphics[height=0.147\textwidth, trim={3.7cm 0 0 0},clip]{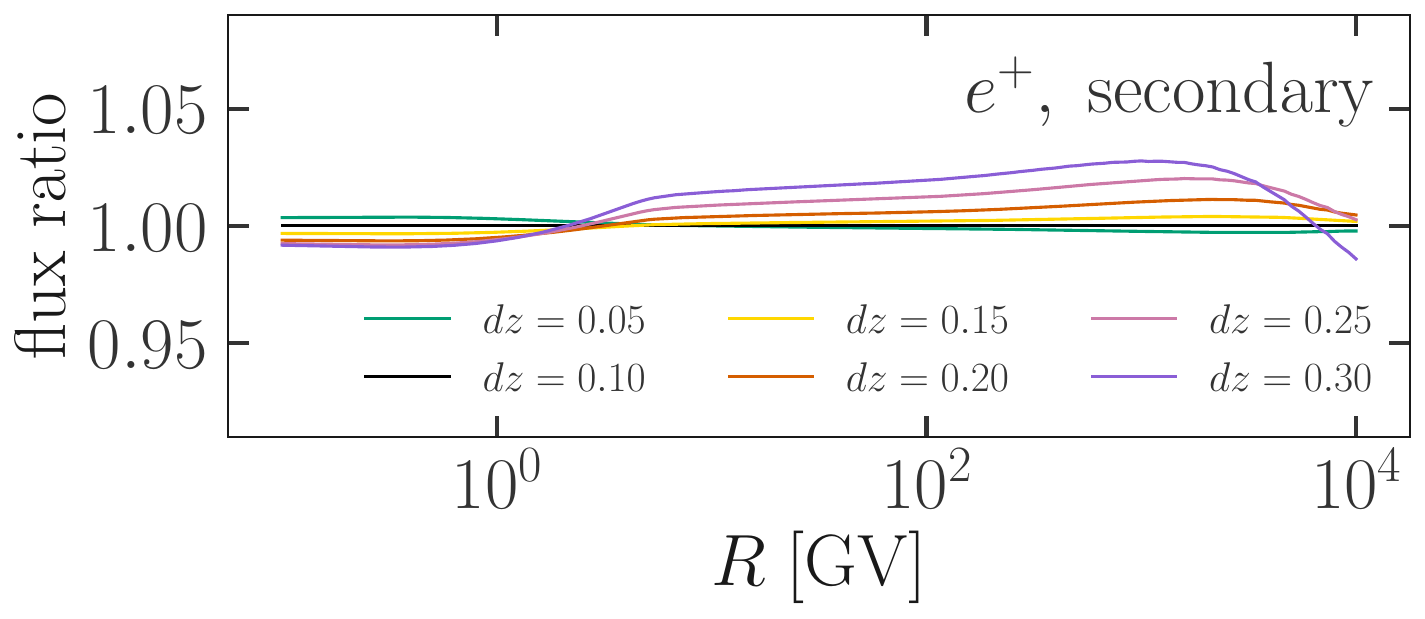}
    \includegraphics[height=0.147\textwidth, trim={3.7cm 0 0 0},clip]{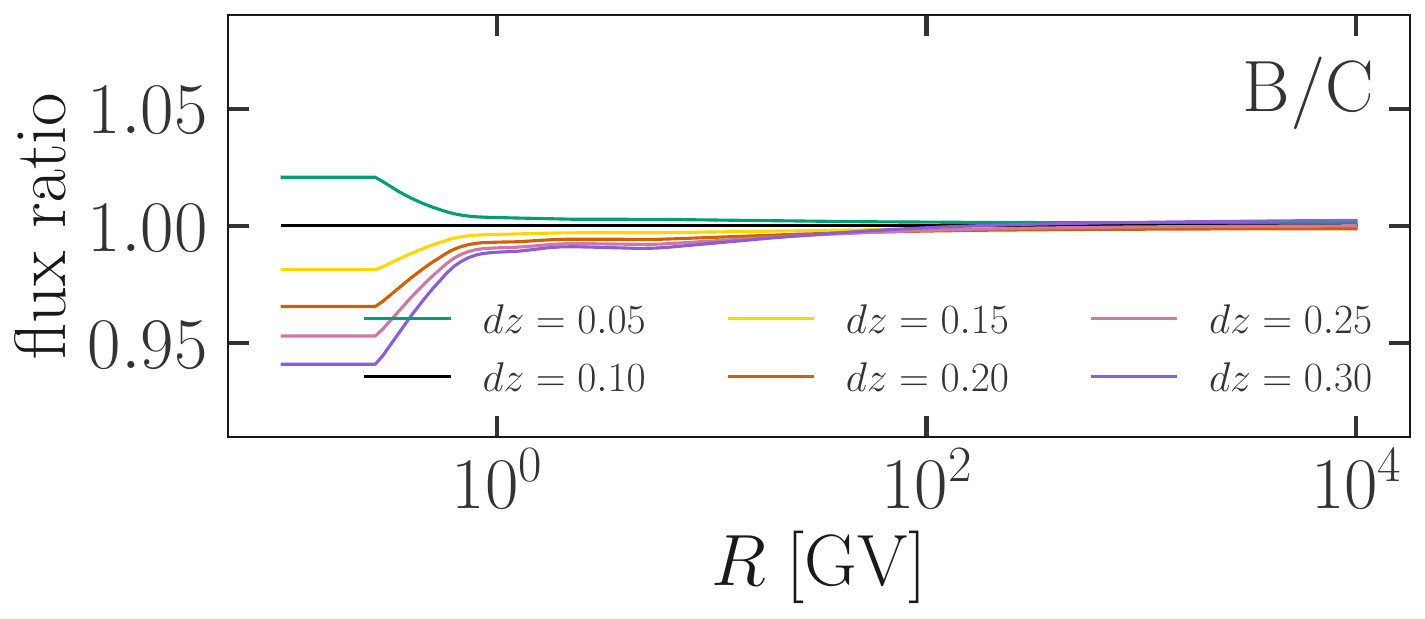}
    \caption{Results of the tests we performed to find the appropriate choice of the grid. In each plot we show the ratio between the different tested cases and our benchmark. The left/central/right panels show the results for the p, $e^+$ and B/C. The different rows, from the top to the bottom, show the tests performed on the $f_{E_{\rm{kin}}}$, time grid, $dr$ and $dz$.}
    \label{fig:grid}
\end{figure*}

We perform dedicated tests to verify that this is the appropriate choice. In particular, we run \textsc{Galprop} with different choices for the space, time and kinetic energy grid. 
We vary the grid in $z$ by choosing $dz$ from 0.05 to 0.30 kpc, in $r$ by varying $dr$ from 0.2 to 1.8, in $E_{\rm kin}$ by using $f_{E_{\rm{kin}}}$ from 1.01 to 1.50. Instead, for the time grid we choose a few different combinations of the starting and ending time and the parameters $f_t$ and $r_t$.
We show the results in Fig.~\ref{fig:grid} where we report the ratio between the flux obtained for proton, positrons and B/C with the different choices of the grid with respect to our benchmark case.
The kinetic factor $f_{E_{\rm kin}}$ impacts significantly the positron flux. In fact, values smaller than 1.15 should be considered to keep the systematics below the few $\%$ level. The impact on CRs and B/C is smaller.
The spatial grid should be taken with $dr\leq 1$ kpc and $dz\leq 0.30$ kpc to minimize the systematics due to the grid. In particular, the grid in $z$ affects the primary CRs and B/C with a minimal amount and at energies where the AMS-02 data are not present (e.g., for B/C below 1 GV). Instead, $dr$ affects significantly the positron flux for which values larger than 1 kpc can produce systematics larger than 5$\%$.
Finally, the time grid can generate a systematic that is basically a normalization factor for B/C. For primary CRs, the variation is very minor while for positrons are relevant only at energies below 1 GeV where the data have large errors.

\end{document}